\documentclass[prl,superscriptaddress,twocolumn]{revtex4-1}

\usepackage{color}
\usepackage[english]{babel}
\usepackage[T1]{fontenc}
\usepackage[latin9]{inputenc}
\usepackage{verbatim}
\usepackage{subfigure}
\usepackage{latexsym}
\usepackage{float}
\usepackage{amsmath}
\usepackage{amsfonts}
\usepackage{graphicx}
\usepackage{times}   
\usepackage{esint}
\usepackage[unicode=true,pdfusetitle,
 bookmarks=false,colorlinks=true,citecolor=blue,urlcolor=blue,linkcolor=red]{hyperref}
\usepackage{ulem}
\makeatletter

\makeatletter

\@ifundefined{textcolor}{}
{%
 \definecolor{BLACK}{gray}{0}
 \definecolor{WHITE}{gray}{1}
 \definecolor{RED}{rgb}{1,0,0}
 \definecolor{GREEN}{rgb}{0,1,0}
 \definecolor{BLUE}{rgb}{0,0,1}
 \definecolor{CYAN}{cmyk}{1,0,0,0}
 \definecolor{MAGENTA}{cmyk}{0,1,0,0}
 \definecolor{YELLOW}{cmyk}{0,0,1,0}
}

\@ifundefined{date}{}{\date{}}
\AtBeginDocument{
  
}
\makeatother

\newcommand{\SAVE}[1]{}

\newcommand{\newt}[1]{{{#1}}}

\begin{document}
\renewcommand\abstractname{}

\title{Determination of Tomonaga-Luttinger parameters for a two-component liquid}
\author{Olabode M. Sule}
\email{sule1@illinois.edu}
\affiliation{Department of Physics, University of Illinois at Urbana-Champaign, Urbana, IL 61801, USA}
\thanks{The first two authors contributed equally to the work}
\author{Hitesh J. Changlani}
\email{hiteshjc@illinois.edu}
\affiliation{Department of Physics, University of Illinois at Urbana-Champaign, Urbana, IL 61801, USA}
\author{Isao Maruyama}
\affiliation{Faculty of Information Engineering, Fukuoka Institute of Technology, 3-30-1 Wajiro, Higashi, Higashi-ku, Fukuoka, 811-0295, Japan}
\author{Shinsei Ryu}
\email{ryuu@illinois.edu}
\affiliation{Department of Physics, University of Illinois at Urbana-Champaign, Urbana, IL 61801, USA}

\date{\today}

\begin{abstract}
We provide evidence for the mapping of critical spin-1 chains, in particular the 
$\mathrm{SU}(3)$ symmetric bilinear-biquadratic model with additional interactions, to free boson 
theories using exact diagonalization and the density matrix renormalization group algorithm. 
Using the correspondence with a conformal field theory with central charge $c=2$,
we determine the analytic formulae for the scaling dimensions in terms of four
Tomonaga-Luttinger liquid parameters. By matching the lowest scaling dimensions, 
we numerically calculate these field-theoretic 
parameters and track their evolution as a function of the parameters 
of the lattice model.
\end{abstract}

\maketitle
\section{Introduction}
Given a strongly correlated quantum system, an important step towards understanding it is to determine its basic properties, 
such as the presence or absence of a gap, the presence of spontaneous symmetry breaking, etc. 
One then asks for more specific information, and ultimately, the complete description of the underlying low-energy physics. 
Quite often, this characterization involves determining an effective field theory. 
Examples include topological field theories describing the full braiding statistics in gapped quantum systems,  
and conformal field theories (CFT), describing the set of independent critical exponents in gapless systems. 
\newt{ Obtaining these conformal exponents is important 
because close to the critical point, the power law behavior of physical quantities 
like magnetic susceptibility is governed by them. These dimensions complement 
the knowledge of the central charge, denoted by $c$, in determining the universal long-distance behavior 
of the theory.}

\newt{In recent times, several probes, such as the entanglement entropy (EE), Renyi entropies 
and entanglement spectrum~\cite{Calabrese_Cardy, Li_Haldane, Ryu, Headrick, 
Tubman, Melko, Ryu_Hatsugai, Thomale2010, Lundgren2014}, have been devised to explore 
the above mentioned properties. A central component of all these measures 
is the ground state reduced density matrix, calculated for a finite region of space 
and obtained by tracing the full density matrix over the other degrees of freedom. 
For example, the finite-size scaling of the EE in one-dimensional critical systems 
provides a precise estimate of the central charge of the corresponding CFT. 
More sophisticated ways of using reduced density matrices also reveal information 
about the low-energy scaling dimensions and operators~\cite{Cheong_Henley, Muender, 
Henley_Changlani, Melko_mutual, Barcza, Furukawa2009}. 
} 

\newt{ For one dimensional (1D) critical systems, the theory of Tomonaga-Luttinger 
liquids (TLLs)~\cite{Tomonaga, Luttinger_original, Giamarchi, Haldane_Luttinger} 
has been remarkably successful at characterizing their low-energy physics. 
There has been additional validation on the experimental front, at least qualitatively; 
several realizations, ranging from carbon nanotubes~\cite{Bockrath, Ishii} 
to semiconductor wires~\cite{Yacoby}, of TLL physics have been found. 
Quantitative estimates of the scaling dimensions, velocity, and Luttinger parameter 
for model Hamiltonians have been made with analytic solutions or numerically, with 
exact diagonalization and density matrix renormalization group~\cite{dmrg_white} 
methods~\cite{Lauchli_operator, Jeckelmann_DMRG, Karrasch, Dalmonte, Furukawa2009}.} 

\begin{figure}[htpb]
\centering
\centering
\includegraphics[width=\linewidth]{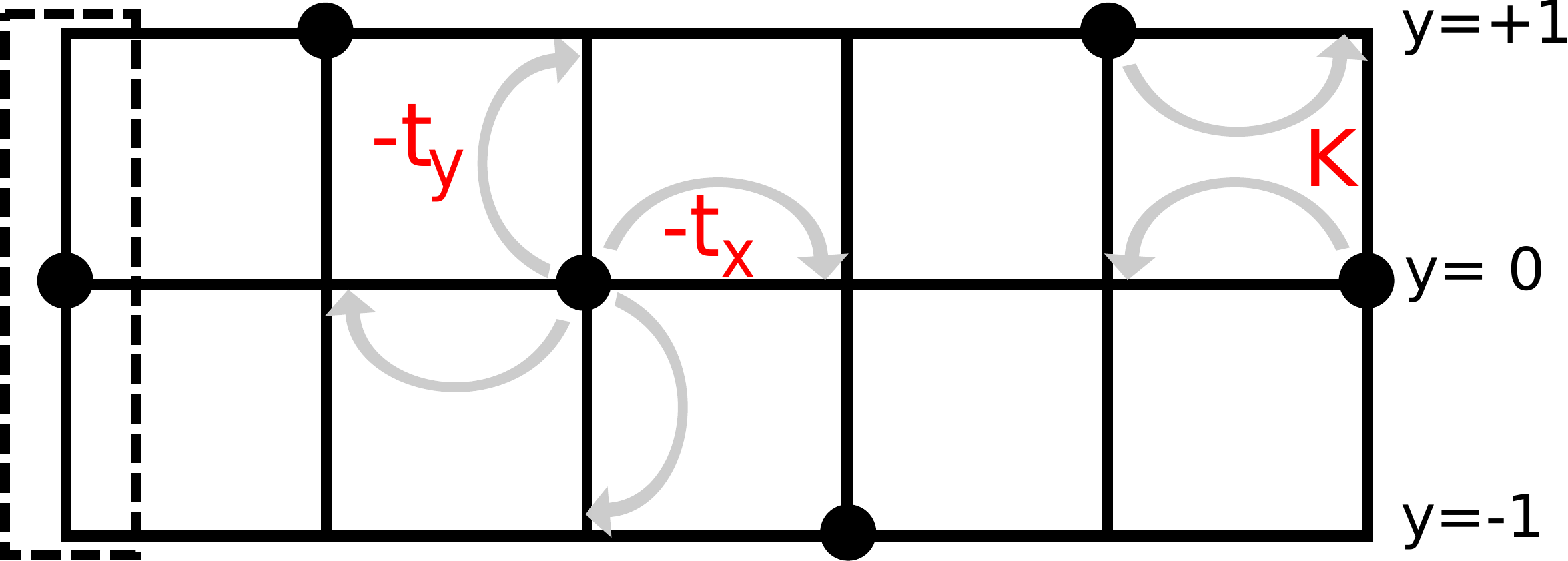}
\label{fig:three_leg_ladder}
\caption{A representative configuration of hard-core bosons on 
a three leg ladder with periodic boundary conditions in the transverse or rung 
direction, with Hamiltonian given by Eq.~\eqref{eq:boson_ham}. 
The hopping in the length ($t_x$) and transverse ($t_y$) directions respectively 
and the correlated hop ($K$) on the square plaquette have been indicated by arrows. 
For $t_x = 0$ and at 1/3 filling, the low energy model involves configurations with 
only with exactly one boson per rung. The three configurations per rung, 
one of which has been enclosed in a dotted rectangle, when appropriately 
Fourier transformed, are mapped to a spin-1 basis. The resultant spin-1 Hamiltonian has 
the form Eq.~\eqref{eq:Hamiltonian}}
\end{figure}	

Most theoretical studies have focused on the single component TLL, which directly 
corresponds to a $c=1$ CFT, and which now appears to be a fairly well 
understood case~\cite{Giamarchi,Furukawa2009}. In contrast, there are 
few general results for the $c=2$ case, despite the existence of systems with 
this property~\cite{Rex_TLL,Sakai}. This is partly attributed to the TLL theory for $c=1$ being 
completely described by a single dimensionless parameter, whereas the $c=2$ theory requires 
four dimensionless parameters. \newt{An important open question is that
there is no established method to extract TLL parameters 
for a given lattice model.} 
Given the history of the TLL, it appears to us that this situation 
is quite unsatisfactory and incomplete. 

In special cases, a $c=2$ CFT can be understood as a tensor 
product of two $c=1$ CFTs; for example, the 
1D Hubbard model has two TLL parameters, one for spin and the other for charge. 
In this paper, however, we will demonstrate a TLL parameter extraction procedure 
for a $c=2$ CFT where such a decomposition does not apply. 
\newt{Several conceptual and practical questions arise here; including 
which measures must be calculated to estimate 
them and whether they are numerically 
accurate enough to validate or refute a given field theory. 
Our paper addresses these questions and highlights an interesting application of 
relatively new ground state entanglement based metrics, 
such as the mutual information. However, before considering a specific problem to demonstrate 
our ideas for $c=2$ CFTs, we mention physical examples where this situation occurs. 

One way to realize a multi-component TLL is to couple several TLLs~\cite{Schulz1, Schulz2}.
The most natural geometry for doing this is a ladder (or tube), a quasi-one dimensional system 
made up of one dimensional legs which are additionally coupled in the transverse or rung direction, 
with open (or periodic) boundary conditions. Fig.~\ref{fig:three_leg_ladder} 
shows an example with three legs, relevant for modelling 
quasi one-dimensional compounds such as [(CuCl$_2$tachH)$_3$Cl]Cl$_2$~\cite{Schnack} 
and CsCrF$_4$~\cite{Manaka}, and to which recent theoretical 
works~\cite{Bose_metal, Balents_spin_tube, Sakai, Sato_PRB} have been devoted.}

\newt{Following the work of Ref.~\cite{Bose_metal} and as is 
schematically depicted in Fig.~\eqref{fig:three_leg_ladder}, our starting point is a 
system of hard-core bosons on a three leg tube, governed by the Hamiltonian,
\begin{subequations}
\begin{eqnarray}
	H &=& H_{\text{hop}} + H_K  \label{eq:boson_ham} \\
	H_{\text{hop}} &=& -t_x \sum_{{\bf r}} b_{{\bf r}}^{\dagger} b_{{\bf r}+  \hat{x} } + \text{h.c.} - t_y \sum_{{\bf r}} b_{{\bf r}}^{\dagger} b_{{\bf r}+\hat{y}}  + \text{h.c.} \\
	H_K &=& K  \sum_{{\bf r}} b_{{\bf r}}^{\dagger} b_{{\bf r}+  \hat{x} }  b_{{\bf r}+  \hat{x} + \hat{y} }^{\dagger}  b_{{\bf r}+  \hat{y} }    + \text{h.c.}
\end{eqnarray}
\end{subequations}
where $t_y$ and $t_x$ are the hoppings along the transverse ($y$) and length ($x$) 
directions respectively, $K$ is a correlated exchange on a square plaquette. 
The phase diagram of this model is expected to be quite rich; here 
we only consider the case of $t_x, t_y \rightarrow 0 $ with 1/3 filling of bosons. In this 
parameter regime, the low-energy theory of this model involves 
only one boson per rung (column) allowing three distinct states on it; 
the number per rung can not change because of the absence of hopping 
in the $x$ direction. 

On Fourier transforming bosonic creation operators along the $y$ direction, 
a new basis at every $x$ location is defined as, 
$| 0 \rangle_x \equiv \frac{1}{\sqrt{3}} \left( b_{x,0}^{\dagger} + b_{x,1}^{\dagger} + b_{x,-1}^{\dagger} \right) |vac\rangle $ 
and $| \pm \rangle_x \equiv \frac{1}{\sqrt{3}} \left( b_{x,0}^{\dagger} + \omega b_{x,\pm 1}^{\dagger} 
+ \omega^{2} b_ {x,\mp 1}^{\dagger} \right) |vac\rangle$ where $\omega=\exp \left( i 2\pi/3 \right) $. 
The three components can be thought of as those corresponding to a pseudospin-1 object, 
leading to the effective spin-Hamiltonian of the form~\cite{Bose_metal},}
\begin{subequations}
\begin{eqnarray} \label{eq:Hamiltonian}
{H} &=& K \Big( \sum_{\langle ij\rangle}{\mathbf{S}_{i}\cdot{\mathbf{S}_{j}}} + \sum_{\langle ij\rangle} \left({\mathbf{S}_{i}} \cdot{\mathbf{S}_{j}} \right)^{2} \Big) + Q_{x} + Q_{y}\\
	Q_{x(y)}&=& q_{x(y)} \sum_{i} {U_{x(y)}^{\dagger i} U^{i+1}_{x(y)} } + \text{h.c.} \label{eq:Bose_metal}
\end{eqnarray}
\end{subequations}
where $\mathbf{S}_i$ is a spin-1 operator living on site $i$, 
while $q_x$ and $q_y$ are scalars. \newt{ $K$ will be set to $1$ throughout 
and thus all energy scales in this paper are in terms of this unit.} 
$U_x$ and $U_y$ are $3 \times 3$ matrices 
in the $S_z$ basis (ordered as $-1,0,1$), and are given by, 
\begin{eqnarray}
U_x =
\left(
\begin{array}{ccc}
\omega^{-1}  & 0 & 0 \\
0 & 1 & 0 \\
0 & 0 & \omega
\end{array}
\right),
\quad
U_y
=
\left(
\begin{array}{ccc}
0 & 0 & 1 \\
1 & 0 & 0 \\
0 & 1 & 0
\end{array}
\right),
\end{eqnarray}

\newt{The $Q_y$ term is non-zero with $q_y = -1/3$, but 
there is no $Q_x$ term i.e. $q_x=0$. Physically, the term $U^{i,\dagger}_y U^{j}_y$ 
models a correlated cyclic permutation of neighboring spins. 
However, diagonalizing $U_y$, i.e. performing a 
similarity transformation by the matrix, 
\begin{eqnarray}
S =
\frac{1} {\sqrt{3}}\left(
\begin{array}{ccc}
\omega^{-1}  & \omega & 1 \\
1 & 1 & 1 \\
\omega & \omega^{-1} & 1
\end{array}
\right),
\end{eqnarray}
preserves the combined $\mathbb{SU}(3)$ symmetry of the first 
two terms in~\eqref{eq:Hamiltonian} and converts the $Q_y$ term 
into the $Q_x$ term because $S U_y S^{\dagger} = U_x$.  
For presentational purposes, we have shown both terms in Eq.~\eqref{eq:Hamiltonian}; 
this generalized model has been previously introduced in the literature 
as the quantum torus chain~\cite{Qin_torus}.}

For $q_x=q_y=0$, this model is the analytically solvable Lai-Sutherland 
model~\cite{Lai,Sutherland,Itoi_Kato}, which serves as a 
useful guide for checking our calculations. Since $Q_x$ and $Q_y$ 
are related by a $\mathrm{SU(3)}$ unitary transformation; 
studying the model with $q_x$ non zero and $q_y=0$ is 
equivalent to the case with $q_x=0$ and $q_y$ non-zero. 
We set $q_y=0$ throughout this paper, and leave the more general 
case for later exploration. Finally, we note that a 
generalized version is the bilinear-biquadratic model~\cite{Papa}, 
whose phase-diagram includes a gapless phase and the gapped Haldane phase~\cite{Haldane,AKLT} 
and which has been experimentally realized in LiVGe$_{2}$O$_{6}$~\cite{Millet_expt}.

We now discuss the organization of the remainder of the paper. 
In Sec.~\ref{sec:bosonization}, we discuss how the 
low-energy theory of the spin-1 model~\eqref{eq:Hamiltonian}, motivated above, is 
mapped to a field theory using bosonization techniques. 
We then develop the analytic formulas 
for the scaling dimensions of the low-energy theory in terms 
of the TLL parameters: these formulae are generalizations 
of those known for the $c=1$ case~\cite{Oshikawa_TLL}. For the particular 
case of parameters of the spin-1 Hamiltonian~\eqref{eq:Hamiltonian} ($q_x > 0$, $q_y=0$), 
these formulae show the explicit dependence of the TLL parameters on 
the microscopic model parameter. In Sec.~\ref{sec:numerics}, we provide numerical 
evidence for the connection between the low energy theory of the spin chain 
and the CFT for $c=2$, by calculating 
the lowest two scaling dimensions with 
exact diagonalization (ED) and the density matrix renormalization group (DMRG). 
The TLL parameters obtained are tracked as a function of the microscopic parameter $q_x$. 
Finally in Sec.~\ref{sec:conclusion}, we conclude by discussing generalizations 
of our method and the prospective applications to other systems.

\section{Mapping Spin-1 Lattice models to free boson theory}
\label{sec:bosonization}

\subsection{Symmetries}
\label{sec:symmetries}
In this section, we develop a continuum field theory description for the lattice Hamiltonian (\ref{eq:Hamiltonian}), 
by closely following Refs.~\cite{Affleck_Les,Itoi_Kato}, wherein more details are spelled out.
For a start, symmetry properties of the Hamiltonian (\ref{eq:Hamiltonian}) are described here piece by piece.  
To this end, the spin-1 part of the Hamiltonian (\ref{eq:Hamiltonian}) can be written (up to a constant factor)
in a manifestly $\mathrm{SU(3)}$ symmetric way 
in terms of $3\times 3$ elementary matrices $L^{\alpha}_{\beta}$ 
with one on row $\alpha$ and column $\beta$ and zero everywhere else as
\begin{align}
{H}_{ \text{SU(3)}} = \sum_{\langle ij\rangle} \sum_{\alpha, \beta=0,1,2} L^{\alpha}_{\,\beta \, i}L^{\beta }_{\,\alpha \, j}.
\label{Lai-Sutherland model}
\end{align}
For convenience, this Hamiltonian can be represented in terms of fermionic operators 
using $L^{\alpha}_{\,\beta \, i} = c^{\alpha\dagger}_i c_{\beta\, i}$, 
with the constraint 
$
\sum_{\alpha=0,1,2} c^{\alpha\dagger}_i  c_{\alpha\, i} 
=
\sum_{\alpha=0,1,2} n_{\alpha i} 
= 1
$
at each site $i$. 
The constraint ensures that the operators have the same commutation (anticommutation) 
relations and act on Hilbert spaces of the same dimensions. 
The Hamiltonian $H_{\mathrm{SU(3)}}$
conserves 
the particle numbers
$N_1-N_0$ 
and
$N_2-N_0$,
where 
$
N_{\alpha} = \sum_i c^{\alpha \dag}_i c_{\alpha i}
$. 
Defining the dual basis by 
$
\tilde{c}_{n} = {3}^{-1/2} \sum_{\alpha=0}^2 {c}_{\alpha} w^{n} 
$
for $n=0,1,2$, 
and the corresponding particle numbers as 
$
\tilde{N}_{n} = \sum_i \tilde{c}^{n \dag}_i \tilde{c}_{n i}
$,
the Hamiltonian $H_{\mathrm{SU(3)}}$
also conserves 
the dual particle numbers
$\tilde{N}_1-\tilde{N}_0$ 
and
$\tilde{N}_2-\tilde{N}_0$.

On the other hand, the $Q_x$ perturbation in (\ref{eq:Hamiltonian}) 
can be written as 
\begin{align}
Q_x = 3 q_x \sum_{i, \alpha}
L^\alpha_{\alpha \,i} L^\alpha_{\alpha \, i+1}.
\end{align} 
The Hamiltonian $H_{\mathrm{SU(3)}}+Q_x$
conserves 
the particle numbers
$N_1-N_0$ 
and
$N_2-N_0$,
and it conserves the dual particle numbers
$\tilde{N}_1-\tilde{N}_0$ 
and
$\tilde{N}_2-\tilde{N}_0$
(mod 3).

\subsection{Continuum theory}

The low-energy effective field theory for the Hamiltonian $H_{\mathrm{SU(3)}}$ can be developed 
by noting that at low energies, only excitations close to the Fermi points 
$k_F = \pi /3a_0$ (where $a_0$ is the lattice constant) propagate. 
Thus we can approximate, 
\begin{align}\label{approx}
c_{\alpha \,i} \approx \sqrt {a_0} [e^{ik_Fx_i }\psi_{R\alpha}(x_i) + e^{-ik_Fx_i }\psi_{L\alpha}(x_i) ].
\end{align}
Substituting this in the Hamiltonian and dropping oscillatory terms, 
the low energy theory can be written in terms of the $U(3)$ currents, 
\begin{align}
J_{R,L \,\beta}^{\alpha} = \psi_{R,L}^{\alpha \dagger} \psi_{R,L \,\beta},
\end{align}
as
\begin{align}
&
{H}_{ \text{SU(3)}} = \pi v_F \int dx \sum_{\alpha, \beta} [J_{R \,\beta}^{\alpha} J_{R \,\alpha}^{\beta}+J_{L \,\beta}^{\alpha} J_{L \,\alpha}^{\beta} + 2J_{R \,\beta}^{\alpha}J_{L \,\alpha}^{\beta}\nonumber\\
&\qquad
-2\cos(2k_Fa_0)J_{R \,\alpha}^{\alpha}J_{L \,\beta}^{\beta} ],
\end{align}
where $v_F$ is the Fermi velocity, which will be set to 1 henceforth. 
The last term depends only on the charged modes, 
$J^{\alpha}_{R\alpha}$ and $J^{\alpha}_{L\alpha}$, 
which are gapped,
while the second term can be shown to be marginally irrelevant in the RG sense. 
\newt{While this term must be retained to evaluate quantitative finite size logarithmic 
corrections, here we simplify the analysis by working directly in the conformal limit. Instead, 
the finite size corrections will be reintroduced only at a later stage, 
when comparing the analytic results with numerical calculations. 
Thus, with this simplification, the critical theory is,}
\begin{align}
{H}_{ \text{SU(3)}} \approx {H}_{\mathrm{WZW}} =  
\pi \int dx \sum_{\alpha, \beta} [J_{R \,\beta}^{\alpha} J_{R \,\alpha}^{\beta}+J_{L \,\beta}^{\alpha} J_{L \,\alpha}^{\beta} ].
\end{align}
We note that this is a $U(3)$ Wess-Zumino-Witten~(WZW) model and thus contains an $\mathrm{SU(3)}_1$ 
and $U(1)$ part~\cite{yellowbook}. 
The $U(1)$ piece is precisely the charged mode which 
is gapped and will be dropped later.
(N.B. the above procedure is better described and applied, 
instead of dealing with the $\mathrm{SU(3)}$ Lai-Sutherland model,
by starting with the Hubbard type model
$
H = -t \sum_{\langle ij \rangle \alpha}
[
c^{\dag}_{i \alpha} c^{\ }_{j\alpha} + h.c.
]
+
U \sum_{i, \alpha\neq \beta} n_{i\alpha}n_{i\beta}
$
without constraint $\sum_{\alpha} n_{\alpha i}=1$. 
This constraint is in fact generated dynamically and 
this model reduces to the $\mathrm{SU(3)}$ symmetric spin model when expanded in 
$t/U$.) 

Applying the same reasoning as above one deduces the continuum approximation
\begin{align}
\label{qx}
Q_x \approx 3q_x \int dx \sum_{\alpha} \left[ (J_{R \,\alpha}^{\alpha})^2+( J_{L \,\alpha}^{\alpha} )^2 \right],
\end{align}
where again we have dropped the terms that only depend on the charged mode. 

\subsection{Abelian Bosonization} 
\label{sec:non_abelian}
Introducing holomorphic and antiholomorphic coordinates for 1+1 d space-time
$z = -i(x-t)$, 
and 
$\bar{z} = i(x+t)$,
the time evolution of the fields factorize nicely so that the fields with an $R(L)$ subscript depend only on $z(\bar z)$ respectively.
The continuum Fermi fields in Eq.\ (\ref{approx}) can be bosonized as follows
\begin{align} \label{bformula}
\psi^{\ }_{\beta L} = \frac{1}{2\pi  a_0} :e^{-i \sqrt{4\pi} \phi_{\beta L}}:, 
\quad
\psi^{\beta \dagger}_L = \frac{1}{2\pi a_0} :e^{i \sqrt{4\pi} \phi_ {\beta L}}:,
\end{align}
where $\phi_L$ represents the holomorphic part of a free boson field. 
We focus on the holomorphic parts of the theory (dropping the L subscript) with similar formulae for 
left moving fermions in terms of the anti-holomorphic part of the free boson field.  
We have introduced normal ordering of an operator $O$, denoted by $:O:$, which must be used when 
two fields at the same point are multiplied together. Usually when bosonizing more that one species of fermions, 
one introduces Klein factors to ensure that different Fermi fields anticommute. 
These Klein factors have been ignored here since they are not dynamical and do not play a role in the Hamiltonian 
which is mainly what we are interested in here. 

A key identity, which can be regarded as the inverse of Eq.\ (\ref{bformula}),
is
\begin{align}
\psi^{\alpha\dagger}\psi^{\ }_{\alpha}(z) = 
\frac{-i}{\sqrt{\pi}} \partial \phi_{\alpha} (z),
\end{align}
where $\partial$ denotes a derivative with respect to $z$.
To understand the split into $\mathrm{SU(3)}$ and $\mathrm{U(1)}$ WZW theories mentioned above 
we introduce the $\mathrm{SU(3)}$ and $\mathrm{U(1)}$ currents given by
\begin{align}
J^a= 
\sum_{\alpha,\beta}
\psi^{\alpha \dagger} T^a_{\alpha \beta}\psi_\beta, 
\quad
J = \sum_{\alpha} \psi^{\alpha \dagger} \psi^{\ }_{\alpha},
\end{align}
where $T^a$ are generators of the $\mathrm{SU(3)}$ algebra. 
The $U(1)$ piece in the boson language satisfies 
\begin{align}
J = \frac{-i}{\sqrt{\pi}}\left(\partial \phi_0 + \partial \phi_1+\partial \phi_2\right). 
\end{align}
The $\mathrm{SU(3)}$ currents associated to the Cartan sub-algebra are
\begin{align}
H^1 &\propto \psi^{0\dagger}\psi_0-\psi^{1\dagger}\psi_1 \propto \partial\phi_0-\partial\phi_1, 
\nonumber\\
H^2 &\propto \psi^{0\dagger}\psi_0+\psi^{1\dagger}\psi_1-2\psi^{2\dagger}\psi_2  \propto\partial\phi_0+\partial\phi_1-2\partial\phi_2.
\end{align}
So we can make an operator product expansion~(OPE) preserving orthogonal change of basis
to introduce $\tilde{\phi}_{0,1,2}$ as
\begin{align}
\tilde \phi_0 &= (\phi_0+\phi_1+\phi_2)/\sqrt{3},
\\ \nonumber
\tilde \phi_1 &= (\phi_0-\phi_1)/\sqrt{2},
\\ \nonumber
\tilde \phi_2 &= (\phi_0 +\phi_1-2\phi_2)/\sqrt{6}.
\end{align} 
In this basis the dynamics of the charged mode is 
now encoded in the single boson field $\tilde \phi_0$. 
Therefore dropping the charged mode corresponds to setting $\tilde \phi_0 = 0$. 
This is indicated with an arrow in the equations below.
In this basis some of the $\mathrm{SU(3)}$ currents associated with the Cartan subalgebra are simply (up to 
a constant factor) $\partial\tilde\phi_1$, $\partial\tilde\phi_2$, 
while those associated with a choice of simple roots for $\mathrm{SU(3)}$ are
\begin{align}
J^{\boldsymbol{\alpha}_1} 
&\propto
\psi_0^\dagger \psi_1 
\propto  e^{i \sqrt{4\pi} (\phi_1-\phi_0)} \rightarrow 
e^{i \sqrt{8\pi} \boldsymbol{\alpha}_1 \cdot \tilde{\boldsymbol{\phi}}},
\nonumber\\
J^{\boldsymbol{\alpha}_2} 
&\propto
\psi_0^\dagger \psi_2 
\propto e^{i \sqrt{4\pi} (\phi_2-\phi_0)}\rightarrow 
e^{i \sqrt{8\pi} \boldsymbol{\alpha}_2 \cdot \tilde{\boldsymbol{\phi}}}
\nonumber
\label{currents}.
\end{align}

$\boldsymbol{\alpha}_1$ and $\boldsymbol{\alpha}_2$  together with a third root $\boldsymbol{\alpha}_3$  are given by,
\begin{align}
 \boldsymbol{\alpha}_1 = (1,0),
 \,
 \boldsymbol{\alpha}_2 = \left(\frac{1}{2}, \frac{\sqrt{3}}{2}\right),
 \,
 \boldsymbol{\alpha}_3 = \left(\frac{1}{2}, -\frac{\sqrt{3}}{2}\right). 
\end{align}
All other roots of $\mathrm{SU(3)}$ can be obtained as integer linear combinations of $\boldsymbol{\alpha}_1$ and $\boldsymbol{\alpha}_2$, for example $\boldsymbol{\alpha}_3= \boldsymbol{\alpha}_1-\boldsymbol{\alpha}_2$. Similarly, 
the vertex operators associated with all other roots can be obtained from operator products of  $J^{\boldsymbol{\alpha}_1}$ and $J^{\boldsymbol{\alpha}_2}$.
\newt{This} construction gives precisely the vertex operators obtained in the purely 
bosonic construction of $\mathrm{SU(3)}_1$ where the boson fields $\tilde \phi_1$ 
and $\tilde \phi_2$ are compactified on the root lattice of the $\mathrm{SU(3)}$ 
algebra~\cite{GSW}. All proportionality constants can be fixed by a choice of normalization of the $\mathrm{SU(3)}$ generators.

We now obtain the purely bosonic description of the gapless degrees of freedom of the spin-1 chain. 
The key identities are
\begin{align}
\sum_\alpha J^{\alpha}_{\alpha} J^{\alpha}_{\alpha} &= -\sum_\alpha \partial \phi_\alpha  \partial \phi_\alpha
\nonumber\\
&\rightarrow -(\partial \tilde \phi_1\partial \tilde \phi_1 + \partial \tilde \phi_2\partial \tilde \phi_2), 
\\
\sum_{\alpha  \neq \beta} J^{\alpha}_{\beta} J^{\beta}_{\alpha} 
&= - \sum_{\alpha \neq \beta} \partial \phi_{\alpha} \partial \phi_{\beta} 
\nonumber\\
& \rightarrow -(\partial \tilde \phi_1\partial \tilde \phi_1 + \partial \tilde \phi_2\partial \tilde \phi_2). 
\end{align}
Using these, we obtain the main results of this section
\begin{align}\label{keyeq}
H_{\mathrm{SU(3)}} \approx -2
\int dx
(\partial \phi_1\partial \phi_1 + \partial \phi_2\partial \phi_2 + \text{antilhol}),\nonumber\\
Q_x \approx -3 q_x 
\int dx
(\partial \phi_1\partial \phi_1 + \partial \phi_2\partial \phi_2 + \text{antilhol}).
\end{align}
where antihol denotes the antiholomorphic part.
Note that the tildes have now been dropped: the effective Hamiltonian of the 
spin-1 chain is now written in terms of $\mathrm{SU(3)}_1$ 
boson fields $\phi_1$ and $\phi_2$. We note that these quantities are all non-negative since we have for any field $\phi$,
\begin{align}
\partial \phi \partial \phi + \bar\partial \phi \bar\partial \phi = -\frac{1}{2}\left(\partial_x\phi \partial_x\phi+ \partial_t \phi \partial_t \phi\right).
\end{align}

\subsection{General $c=2$ Boson theories}
\label{sec:general_boson}

In the previous section, we derived the low-energy effective Hamiltonian that should 
capture the critical dynamics of the spin-1 chain with the $Q_x$ perturbation. 
The low-energy effective theory consists of two compactified boson fields and has the central charge $c=2$.
To put the effective theory in a general context, we discuss in this subsection a generic two-component boson theory with $c=2$.  

For the case of the single-component TLL, 
the landscape of the theory (often called ``moduli space'') is well understood.
It is characterized solely by a single parameter, 
the Luttinger parameter $K$ or the compactification radius $R$ of the boson field.  
There is a boson-vortex duality in (1+1)d (also known as ``T-duality'') which relates 
the two regions $K>1$ and $K<1$. These regions are separated by the self-dual point $K=1$ where
$\mathrm{SU(2)}$ symmetry is realized.
With orbifolding, 
theory space for $c=1$ is described in terms of two axis,
each describing the ordinary free boson theory (the single-component TLL) and its orbifolded counterpart,
together with a few ``exceptional cases''
\cite{Ginsparg_Les}. 

On the other hand, the moduli space for the $c=2$ theories is more complicated.  
For a start, let us consider the action in 1+1 d space-time for two bosonic fields $X^{1,2}$, 
\begin{align}
\label{action2}
S = \frac{1}{4\pi} \int dx dt\,
(G_{a b}\partial_\mu X^a\partial_\mu X^b + B_{a b}\epsilon_{\mu \nu} \partial_\mu X^a \partial_\nu X^b),
\end{align}
where $\mu,\nu=0,1$,
and $G$ and $B$ are a symmetric (non degenerate) and antisymmetric 2 by 2 real matrix, respectively. 
\begin{align}\label{Ghamiltonian}
X^a \sim X^a + 2\pi.
\end{align} 
The corresponding Hamiltonian is given by,
\begin{align}
{H}=-\frac{1}{2\pi } \int dx \, G_{a b}(\partial X^a\partial X^b + \text{antihol}),
\end{align} 
Observe that the parameter $B$ does not enter into the Hamiltonian: 
it is a topological term. However, it affects
the canonical commutation relations and hence the spectrum. 

There are thus four independent parameters, $G_{11}, G_{12}, G_{22}$ and $B_{12}$, characterizing  the $c=2$ action (\ref{action2}), 
as opposed to the $c=1$ TLL parameterized by a single parameter.  
(As in the case of $c=1$, one can consider various orbifolds of 
the two-component boson theory (\ref{action2}), leading to an even richer moduli space
or phase diagram
\cite{Dulat}. ) 

\begin{figure}[htpb]
\centering
\centering
\includegraphics[width=0.8\linewidth]{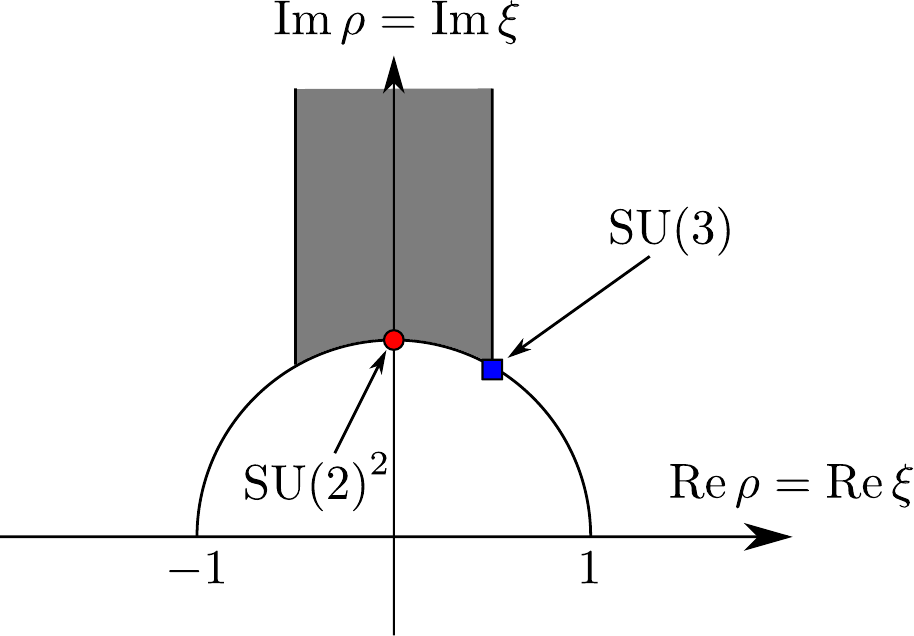}
\caption{
\label{fig:modulis}
{
Theory space of the two-component Tomonaga-Luttinger liquid with $\rho=\xi$.
The shaded region represents the ``fundamental domain'';
Because of the duality, 
different points in the theory space that are related by the duality
are isospectral. 
The fundamental region is a set of representatives for all points related by the duality.
I.e., starting from points in the shaded region, by mapping these points by 
the duality group, the entire theory space is covered. 
Some spectial points in the theory space are also marked:
``SU(3)'' represents the SU(3) WZW theory, 
and ``SU(2)$^2$'' consists of two copies of SU(2) WZW theories,
which may be realized, e.g, as two copies of the XXX spin chain.
}
}
\end{figure}

For the case of $c=1$ TLL, 
the duality relates the large and small compactification radius (the Luttinger parameter). 
Similarly, there is a group of duality transformations 
acting on the four parameters, and 
different values of $G$ and $B$ do not necessarily correspond to different spectra
~\cite{becker2006string, Giveon}.
To unveil this duality group, it is convenient to trade the four real parameters in $G$ and $B$ for two complex parameters $\xi$ and $\rho$ as follows
\begin{align}
\xi  \equiv \frac{G_{12}} {G_{22}} + i \frac{\sqrt{\text{det} \, G}}{G_{22}}, \nonumber \\
\rho \equiv B_{12} + i \sqrt {\det \,G}.
\end{align}
These two parameters can be acted upon by independent $SL(2, \mathbb Z)$ transformations which for $\xi$ is given by
\begin{align}
\xi \rightarrow \frac{a \xi + b}{c \xi + d},
\end{align}
where $a, b, c, d \in \mathbb Z, ad -bc =1$. There is a similar 
independent transformation for $\rho$. 
{
These transformations change the parameters $G$ and $B$ but 
lead to the same spectrum. 
Effectively the target space of the boson fields corresponds to two tori, which are left invariant by 
$SL(2, \mathbb Z) \times SL(2, \mathbb Z)$ transformations. 
There are two further discrete transformations that leave the spectrum invariant:
\begin{align}
(\xi, \rho) \rightarrow (\rho, \xi),
\quad 
(\xi, \rho) \rightarrow (-\bar\rho, -\bar\xi).
\end{align}
When $B_{12} =G_{12} =0$, we have a product of two $c=1$ theories. 
In this case,
the first transformation sends $G \rightarrow G^{-1}$,
which corresponds to two independent duality transformations for each $c=1$ theory.}
Fig.~\ref{fig:modulis} depicts a portion of the space of theories in the $\xi=\rho$ plane together with some points 
of enhanced symmetry. We anticipate that these theories capture the critical behavior of the gapless degrees 
of freedom of spin-1 chains such as the model in Eq.~(\ref{eq:Hamiltonian}).

To deduce the spectrum of these bosonic theories we switch to Euclidean signature 
$t \rightarrow -it$. 
We take space-time to be a torus of modulus $\tau = \tau_1 + i \tau_2$ i.e we compactify Euclidean space-time as $x \sim x + 2\pi$ and $(x,t) \sim (x,t) + (2\pi \tau_1, 2\pi \tau_2) $.  One can quantize using path integrals, the path integral yields a sum over instanton sectors. We can write in each instanton sector
\begin{align}
X_{n,w}^a  = X_{n,w, \,\text{cl}}^a+ X^a_{q}
\end{align}
where 
\begin{align}\label{classical}
 X^a_{n,w,\,\text{cl}}(x,t) = w^a x + \frac{(n^a-w^a\tau_1)t}{\tau_2}
\end{align}
is a classical solution that winds $n$ and $w$ times along the two non trivial cycles on the torus. 
The partition function is
\begin{align}
Z = \sum_{n,w} e^{-S^{cl}_{n,w}}\int [DX_q] e^{-S[X_q]},
\end{align}
where $S^{cl}_{n,w}$ is the classical action evaluated on shell for Eq.\ (\ref{classical}), 
and the quantum path integral is over a continuous uncompactified variable  $X_q$. 
The second term in Eq.\ (\ref{action2}) is a total derivative for periodic functions $X_q$ and can be neglected. 
Thus the integral over $X_q$ yields the determinant of the quadratic differential operator appearing in 
Eq.\ (\ref{action2}) which is just the Laplacian in the spacetime index times the $G$ matrix in the internal index $a$. 
After applying a Poisson resummation in $n$ for the classical contribution one finds
\begin{align}\label{partition2}
Z = \frac{1}{\text{det} G^{1/2}}
\left(\frac{ \tau_2}{\det' \nabla^2}\right)^{c/2} \sum_{p_L, p_R}
e^{2\pi i \tau_1 \left(p\circ p\right)-2\pi \tau_2\left(p\cdot p\right)},
\end{align}
where $c=2$,  
\begin{align}\label{products}
p \circ p &= p_L^T G p_L - p_R^T Gp_R, \nonumber\\
p \cdot p &= p_L^T G p_L + p_R^T Gp_R,
\end{align}
and 
\begin{align}\label{internalmomenta}
p_L &= \frac{1}{2}\left(G^{-1}(n- B w)+ w\right),
\nonumber\\
p_R &= \frac{1}{2}\left(G^{-1}(n- B w)- w\right).
\end{align}
The Laplacian determinant can be regularized as 
$\det' \nabla^2 =   \tau_2 |\eta(\tau)|^{-4}$, where the prime superscript indicates that 
the zero modes have been removed and $\eta(\tau)$ is the Dedekind eta function.

Comparing with the standard formula for a CFT partition function on the torus one can 
read off the spectrum of scaling dimensions 
\begin{align}
\Delta = p_L^T G p_L + p_R^T Gp_R + \sum_{n_L > 0} n_LN^L_n + \sum_{n_R > 0} n_RN^R_n,
\label{eq:scaling_dimensions}
\end{align}
where the last two terms correspond to the determinant of the Laplacian 
and represent, in canonical quantization, harmonic oscillators indexed 
by positive integers $n_{L,R}$. $N^{L, R}_n$ is the occupation number of oscillator $n_{L, R}$. 

\begin{figure}[htpb]
\centering
\centering
\includegraphics[width=0.8\linewidth]{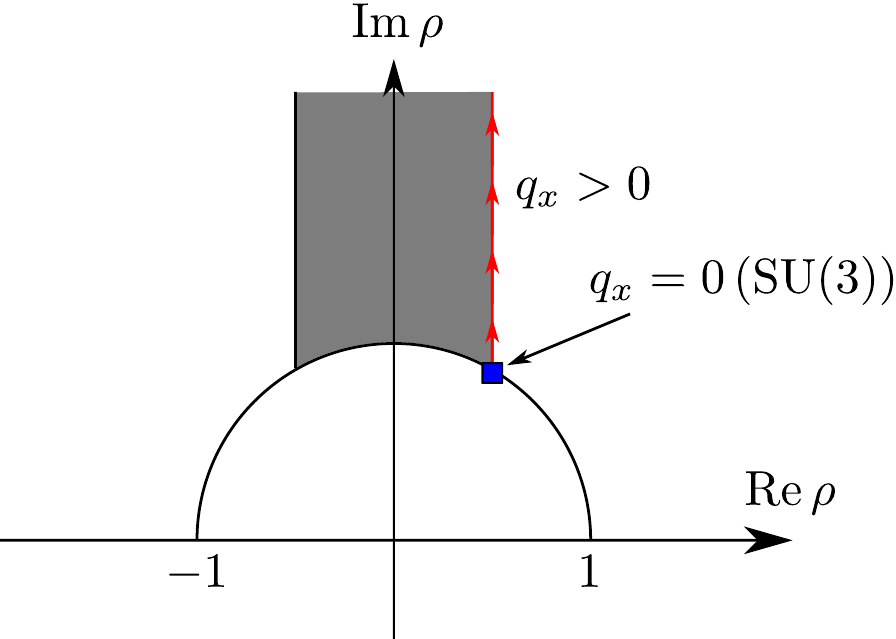}
\caption{
\label{fig:modulis2}
{
Theory space of the two-component Tomonaga-Luttinger liquid with $\xi=1/2+ \sqrt{3}/2 i$.
The point ``$q_x=0 (\mathrm{SU}(3))$'' corresponds to the SU(3) symmetric Lai-Sutherland model.
The red line with arrows represents the points in theory space traversed as 
$q_x$ is increased from 0. (The arrows here do not indicate the renormalization group flow.)}
}
\end{figure}

For the $\mathrm{SU(3)}_1$ WZW theory, 
we take $G$ and $B$ to be,
\begin{align}
 G = \frac{1}{2}\begin{bmatrix} 
 2&{1} \\
  {1} & 2\end{bmatrix},  
 \quad
B = \frac{1}{2}\begin{bmatrix} 0& {1}\\ {-1} & 0\end{bmatrix}.
\end{align}
Here $G$ is proportional to the inverse of the Cartan matrix of $\mathrm{SU(3)}$.
The above choice of parameters at the $\mathrm{SU(3)}$ point is consistent with that used in the numerical sections below. 
The relationship to the fields used in the previous section (note they were tilded) is 
simply a change of basis that diagonalizes $G$ and rescales the diagonal elements to $1$ i.e
\begin{align}
\phi_1 = \frac{1}{\sqrt{8\pi}}(X_1-X_2),
\quad 
\phi_2 = \sqrt{\frac{3}{8\pi}}(X_1+X_2).
\end{align}
Since the terms in Eq.\ (\ref{keyeq}) correspond to the $G$ term in the Hamiltonian (\ref{Ghamiltonian}) 
we deduce that the continuum version of the transformation 
${H}_{\mathrm{SU(3)}} \rightarrow {H}_{\mathrm{SU(3)}} + q_xQ_x$
is
\begin{align}
G_{SU(3)} \rightarrow G_{SU(3)} + q_x\frac{3}{2} G_{SU(3)}.
\label{eq:G_qx}
\end{align}
This prediction will be tested with the help of accurate numerical calculations, 
discussed at length in the next section. Fig. \ref{fig:modulis2}  depicts the portion of the space 
that we traverse starting with our choice of parameters for the $\mathrm{SU(3)}$ 
model and varying $G$ as a function of $q_x$.
The form of the $G$ matrix in Eq.\ (\ref{eq:G_qx})
is consistent with and expected from 
the $\mathbb{Z}_3$ symmetry,
i.e., the conservation of 
$\tilde{N}_1-\tilde{N}_0$
and 
$\tilde{N}_2-\tilde{N}_0$
-- see Sec.\ \ref{sec:symmetries}.
The $\mathbb{Z}_3$ symmetry can be thought of as a $\frac{2\pi}{3}$ rotation in the root 
space of $\mathrm{SU(3)}$. 
In the effective field theory this is represented by the transformation on the currents
$(J^{\boldsymbol{\alpha}_1}, J^{\boldsymbol{\alpha}_2}, J^{\boldsymbol{\alpha}_2})$
(with superscripts defined in
Eq.\ (\ref{currents}) ) as 
$(J^{\boldsymbol{\alpha}_1}, J^{\boldsymbol{\alpha}_2}, J^{\boldsymbol{\alpha}_3})
\to 
(J^{-\boldsymbol{\alpha}_2}, J^{\boldsymbol{\alpha}_3}, J^{-\boldsymbol{\alpha}_1})$. 
In terms of the boson fields $\phi_1$ and $\phi_2$ which live on the $\mathrm{SU(3)}$ root lattice, this amounts to 
\begin{align}
 \boldsymbol{\phi} \to M\cdot \boldsymbol{\phi}, 
 \quad
 M=
 \left(
 \begin{array}{cc}
 -1/2 & \sqrt{3}/2 \\
 -\sqrt{3}/2 & -1/2
 \end{array}
 \right).
\end{align}

In the $\boldsymbol{X}$ basis the $\mathbb{Z}_3$ symmetry is represented by
\begin{align}
 \boldsymbol{X} \to M'\cdot \boldsymbol{X}, 
 \quad
 M' =
 \left(
 \begin{array}{cc}
 0 & 1 \\
 -1 & -1
 \end{array}
 \right).
\end{align}

The $G$ matrix in Eq.\ (\ref{eq:G_qx}) is left invariant under 
the $\mathbb{Z}_3$ transformation. i.e. we have $M'^TGM' = G$. One can show generally that any 
symmetric matrix left invariant by $M'$ is proportional to $G_{SU(3)}$.

\section{Numerical results establishing correspondence of spin chains to conformal field theory}
\label{sec:numerics}
Having described the field theory for $c=2$ spin chains, we now provide 
numerical evidence for the proposed correspondence. 
Our results first focus on various ways of calculating 
scaling dimensions~\eqref{eq:scaling_dimensions}, after which 
we discuss the procedure for extraction of the TLL parameters $G_{ab}$ and $B_{ab}$ for $a,b=1,2$. 
We numerically confirm an important prediction of the field theory, namely Eq.~\eqref{eq:G_qx}.

We carried out ED and DMRG calculations for periodic chains; 
finite size scaling of the energy gaps provides estimates of the lowest scaling dimensions. 
For bigger open chains, we calculate the same information 
from the \emph{mutual information} for spatially disjoint blocks. 
The mutual information measure is completely determined from 
the ground state wavefunction, making it useful for situations where obtaining 
excited states is difficult. 

The numerical calculations in this section were performed with a combination of our own codes and the Algorithms and Libraries for Physics Simulations libraries~\cite{ALPS}.

\subsection {Inferences from Exact Diagonalization and Density Matrix Renormalization Group}
\label{sec:ed_dmrg}
For a 1D periodic chain of length $L$, 
the scaling dimensions $x_j$, corresponding to the $j^{th}$ excited state 
with energy $E_j$, are given by,
\begin{equation}
	E_j - E_0 = \frac{2 \pi v x_j} {L} + \frac{a} {L \log L}
\label{eq:energy_v_scaling_dimension}
\end{equation}
where $a$ is a model specific constant, $v$ is the TLL velocity obtained from 
the finite size scaling of the ground state energy $E_0$,
\begin{equation}
	\frac{E_0}{L} = e_{\infty} - \frac{\pi c v}{6 L^2} + \frac{b} {L^2 (\log L)^{3}}
\label{eq:energy_L}
\end{equation}
where $ e_{\infty}$ is the energy per site in the thermodynamic limit and $c$ is the central charge 
and $b$ is a constant. \newt{The form of the finite size corrections was derived by Itoi and Kato~\cite{Itoi_Kato} for the $\mathbb{SU}(3)$ symmetric point (i.e. $q_x = 0)$; here we have assumed the same form holds for $q_x > 0$.}
 
We note that the above formulas assume all excitations propagate with the same velocity $v$,
while for multi-component TLLs, more than one velocity may appear in general. 
(For more generic models, these formulae need modifications; 
for example see the work of Ref.~\cite{Itoi_Affleck} 
on a $\mathrm{SU(2)} \times \mathrm{SU(2)}$ model.)
In our model, a naive continuum limit and the bosonization analysis, 
(\ref{qx}) and  (\ref{keyeq}), suggests that the excitations of the system,
even when $q_x\neq 0$, should be described by a single velocity. We will take 
this as our working hypothesis. While our spectral analysis by ED/DMRG depends on this assumption,
our \newt{later} analyses based on the entanglement entropy and the mutual information do not. 

\begin{figure}[htpb]
\centering
\centering
\includegraphics[width=\linewidth]{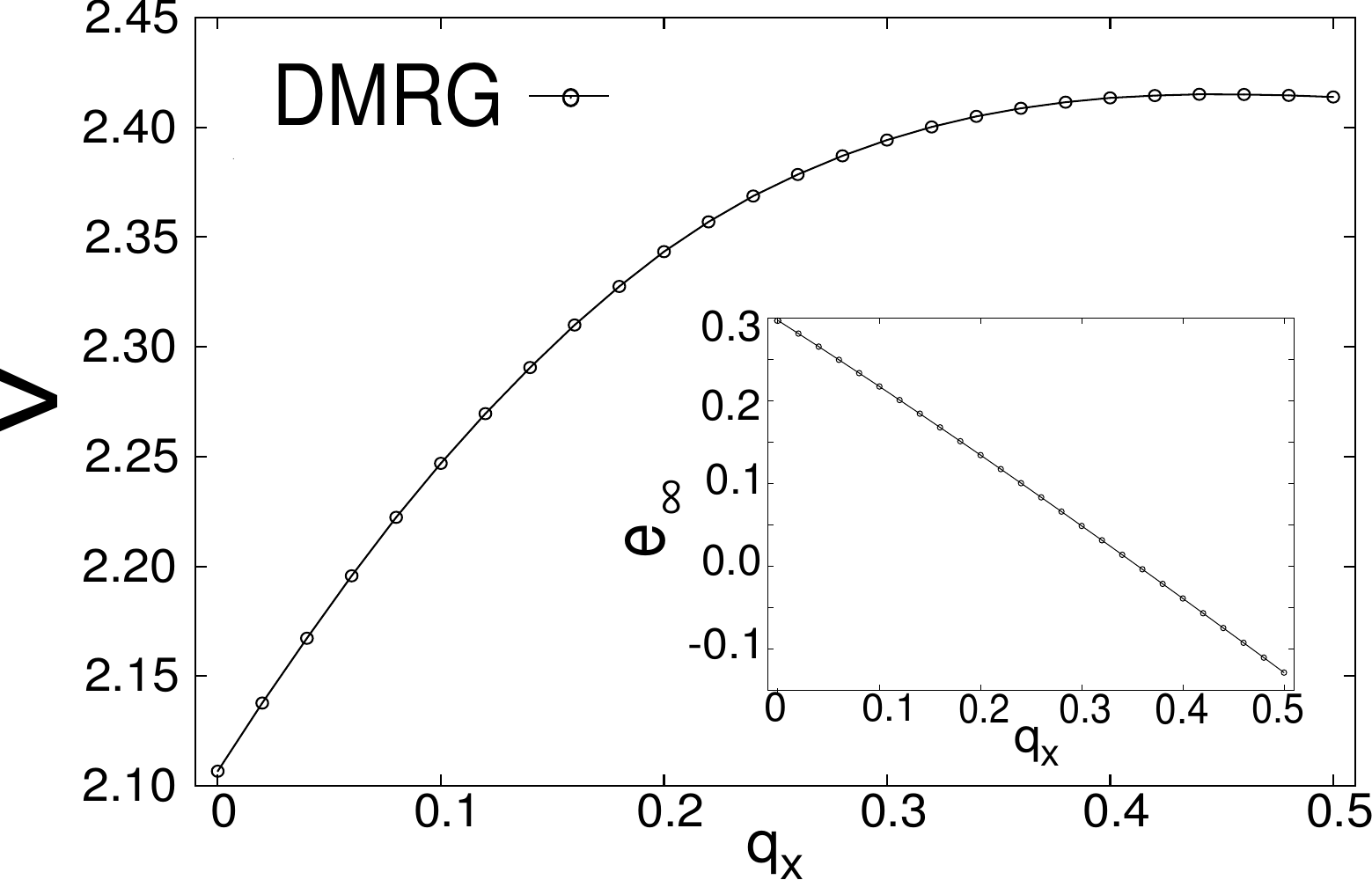}
\caption{
{
Velocity of the coupled TLLs as a function of $q_x$. 
The inset shows the energy per unit length in the thermodynamic limit 
as a function of $q_x$. The lines are guides to the eye.}
}
\label{fig:energy_velocity} 
\end{figure}	

Figure~\ref{fig:energy_velocity} shows the TLL velocity $v$ and ground state energy per site in the thermodynamic limit 
$e_{\infty}$ as a function of $q_x$ obtained by fitting our data to Eq.~\eqref{eq:energy_L}. 
Our results for the $\mathrm{SU(3)}$ symmetric point are in excellent agreement with analytic results~\cite{Fath} 
and previous numerical studies~\cite{Fath,Rachel_SU3,Aguado_SU3,Lauchli_Schmid_SU3}; 
for example, we get  $e_{\infty} = 0.29679$ and $v = 2.107(1)$ which are close to the exact results of 
$2 - \ln 3 - \frac{\pi}{3\sqrt{3}}$  and $2 \pi/3$ respectively. 
Care must be taken in comparing our results with studies which parameterize the bilinear and 
biquadratic terms in the Hamiltonian~\ref{eq:Hamiltonian} to be $J \cos \theta$ 
and $J \sin \theta$ with $\theta = \frac{\pi}{4} $, thus requiring an additional factor of $1/\sqrt{2}$. 
We have used the value of the central charge $c=2$, 
which we established independently from the scaling of the entanglement entropy (EE), discussed next. 

Before we proceed, we mention an important subtlety associated with the 
choice of system sizes used in finite-size scaling. In a previous DMRG study 
on the $\mathrm{SU(3)}$ symmetric model, Ref.~\cite{Aguado_SU3} 
showed the absence of the singlet ground state (scaling dimension 0 in the CFT) 
for chains with lengths $6M+2$ and $6M+4$, where $M$ is a positive integer.~\footnote{This observation can possibly 
be better understood by extracting the scaling operators from numerics. This 
involves determining a coarse grained operator that spans three sites; 
a direction we will not explore in the present paper.}
Thus, we restrict ourselves to analyzing chains with lengths that are multiples of 6.

\subsubsection {Central charge}
\begin{figure}[htpb]
\centering
\centering
\includegraphics[width=1\linewidth]{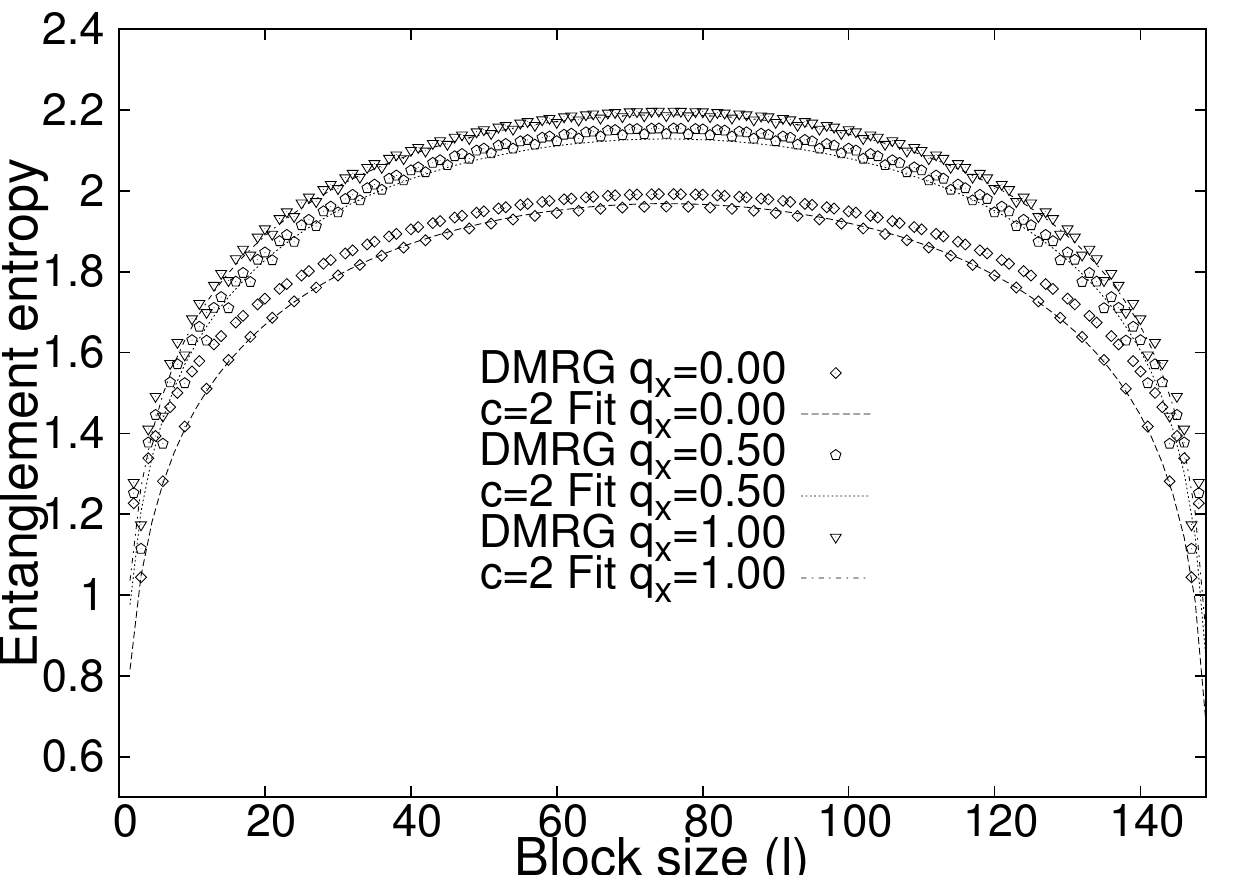}
\caption{Entanglement entropy as a function of block size $(l)$ 
and its fit to the formula~\eqref{eq:cS} for $c=2$ 
for a 150 site open chain for various $q_x$.}
\label{fig:EE} 
\end{figure}

We establish the relevant region in parameter space where the 
TLL physics is expected to hold. For this purpose, we extract the central charge $c$, 
obtained from the scaling of the EE 
of a subsystem or "block", readily available in DMRG, 
as a function of its size $l$. For open chains, the analytic form for the EE, 
denoted by $S(l)$, is,
\begin{equation}
	S(l) = \frac{c}{6} \log \Big ( \frac{L}{\pi} \sin \Big( \frac{\pi l}{L} \Big) \Big) + S_0, 
\label{eq:cS}
\end{equation}
where $S_0$ is a subleading correction. 
In Fig.~\ref{fig:EE} we show the profile of the EE and verify that the 
$c=2$ fit to it is accurate for all $q_x > 0$.~\footnote{Practically this was checked for $0 \leq q_x<100$.} 
However, the EE profile has local structure 
occurring on the scale of three sites, that arise 
due to open boundaries. These are not captured by the leading term in Eq.~\eqref{eq:cS}. 
Other similar quality fits are possible with a lower value of $c$; 
we estimate $c=1.96 \pm 0.05$. Also note that $S_0$ is non-universal; 
in this case dependent on $q_x$ alone. This explains 
why the various curves in Fig.~\ref{fig:EE} differ despite having the same central charge.

We pursue an understanding of the TLL behavior 
for all $q_x > 0 $ by considering the case $q_x \rightarrow \infty$~\cite{Qin_torus}.
In this limit, the model is a purely classical one, 
with a macroscopically large number of ground states. 
To see this, we write out the $Q_x$ term on a bond $\langle i,j\rangle$ 
in terms of $S_z$ and $S_z^2$ operators,
\begin{equation}
	 U_{x}^{\dagger i} U^{j}_{x} + \text{h.c.} = 
	 2-3{S^{i}_{z}}^{2}-3{S^{j}_{z}}^{2} + \frac{9}{2} {S^{i}_{z}}^{2} {S^{j}_{z}}^{2} + \frac{3}{2} S^{i}_{z}S^{j}_{z}. 
\end{equation}
This expression indicates that the configurations $|\mp1 \pm1\rangle$ 
and the configurations $|\pm1 \; 0 \rangle $ (and $|0 \; \pm1 \rangle$) 
are exactly degenerate and have the lowest energies possible. 
This means that starting from a spin-1 "N\'eel" state, for example $|+1 -1 +1 -1 \rangle$,
one can locally replace each $|+1 -1 \rangle$ "dimer" by a $|+1 \; 0 \rangle$ 
without changing the total energy. 
Thus, there is an exponentially 
large number of degenerate states. 
Adding the $\mathrm{SU(3)}$ symmetric term lifts this degeneracy, 
but the model stays critical. Such a macroscopic degeneracy does not exist in 
the spin-1/2 XXZ model in the Ising limit; 
this is why there is a finite value of anisotropy at which the 
spin-1/2 XXZ model ceases to be critical.

\subsubsection{Scaling dimensions and degeneracies}

In order to obtain multiple excited states in the same symmetry sector 
(here sectors of definite $S_z$), we perform a state averaging procedure with two target states 
in the finite system DMRG method. A sequence of bond dimensions 
varying from $m=400$ to $m=2000$ states and periodic chains of lengths 
varying from 24 to 66 sites, were studied. For the ED 
calculations (from 6 to 18 sites), multiple excited states were calculated 
to give us a picture of the low energy degeneracy structure of this model. 

A note about boundary conditions is now in order. Working with open boundary conditions, 
favorable for DMRG, can complicate the mapping of a spin chain to a conformal 
field theory: the notion of strict "conformal invariance" is broken.  
Hence we do not rely on open boundary conditions to give us a picture of the degeneracy 
structure of this system. That said, scaling dimensions 
can still be reliably numerically estimated from open chains. 

For the $\mathrm{SU(3)}$ symmetric model, it is analytically known that 
the first excited state is $18$-fold degenerate in the "conformal limit" 
and the second excited state is $16$-fold degenerate. 
However, in finite size simulations, the conformal limit is reached rather 
slowly as a function of system size; \newt{more specifically} 
the lattice model (\ref{Lai-Sutherland model}) flows into the $\mathrm{SU(3)}_1$ 
WZW critical point only logarithmically fast. Thus, we rely only on \emph{trends} 
seen in the ED results. 

\begin{figure*}[htpb]
\centering
\subfigure[]{\includegraphics[width=0.31\linewidth]{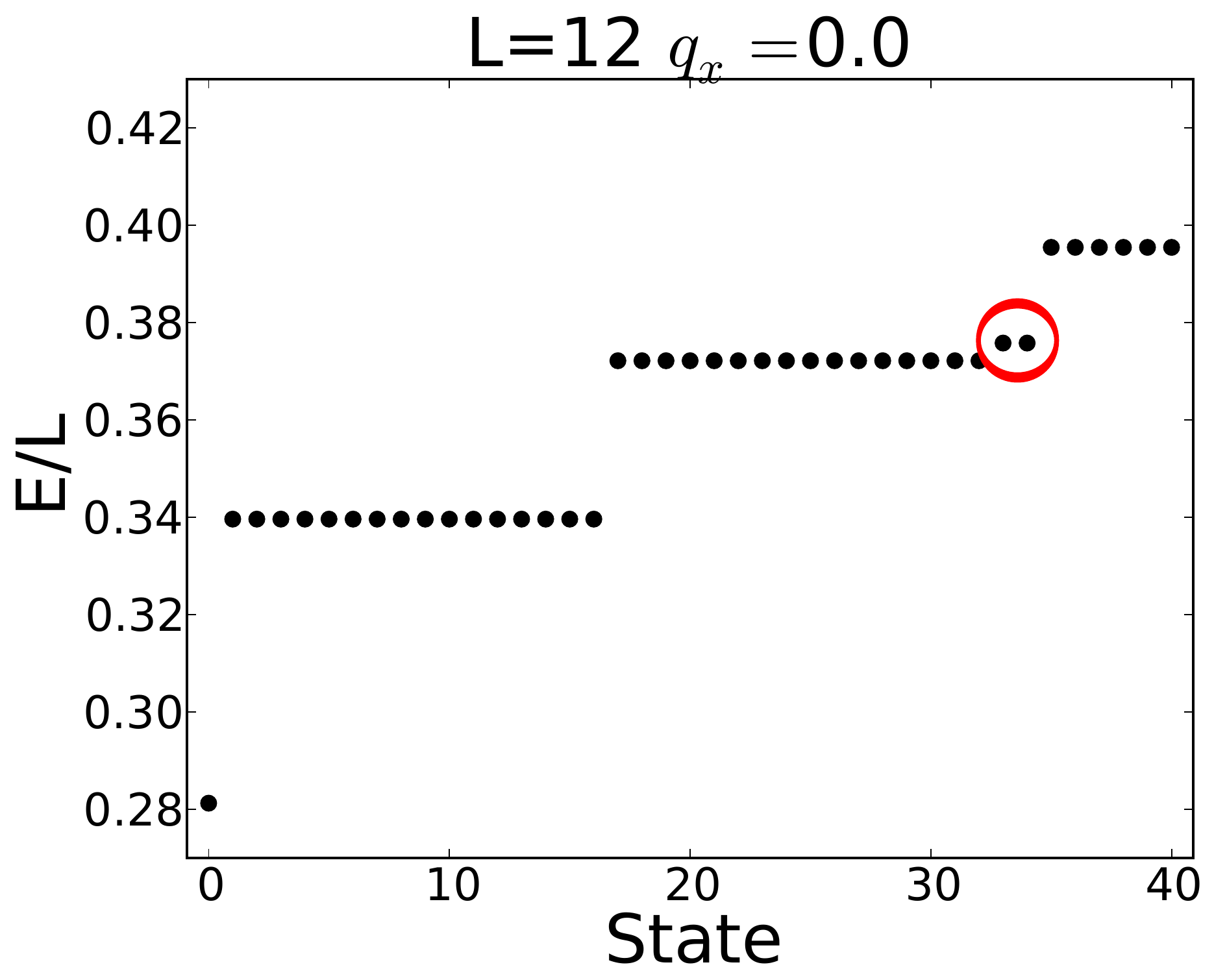}}
\subfigure[]{\includegraphics[width=0.31\linewidth]{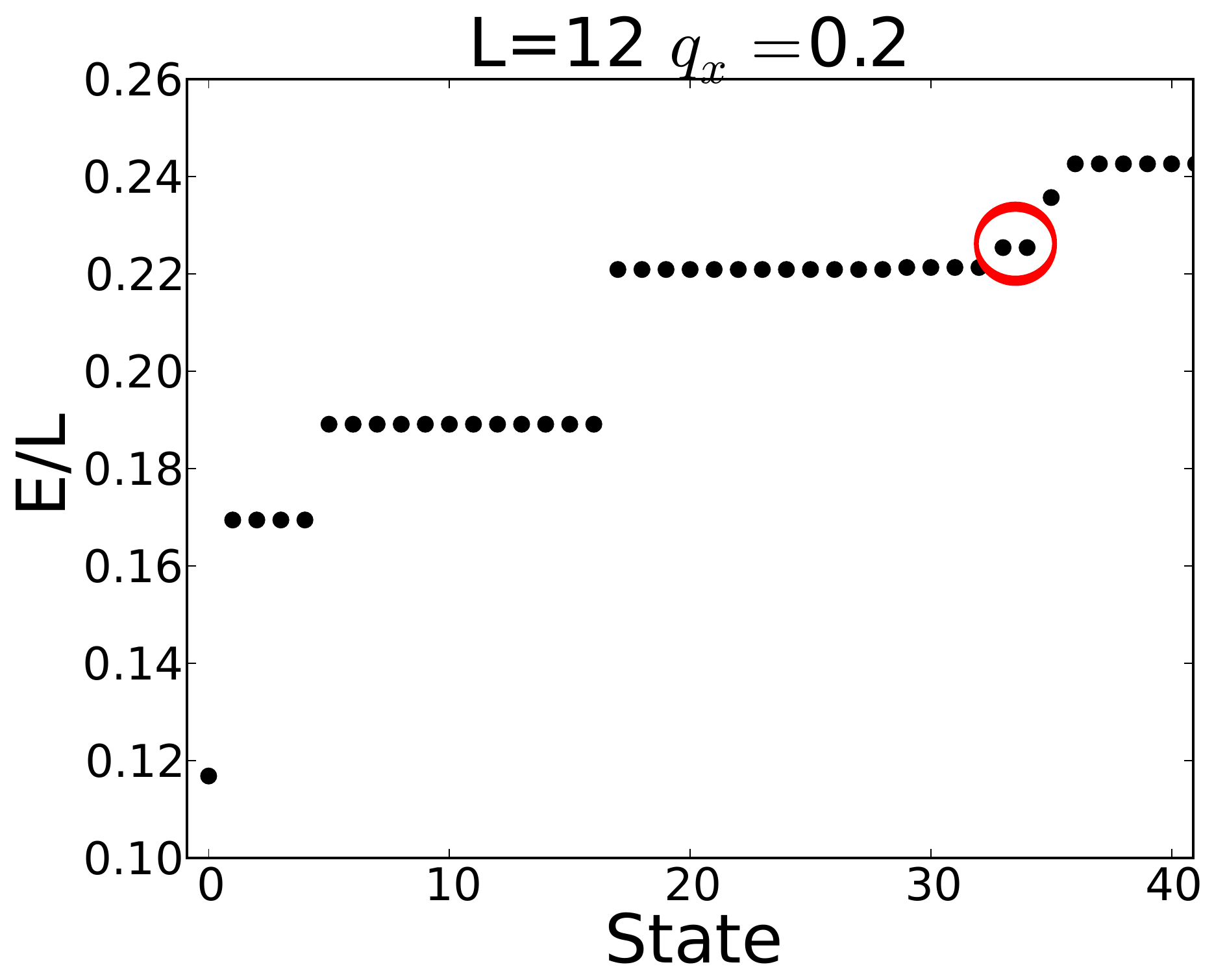}}
\subfigure[]{\includegraphics[width=0.32\linewidth]{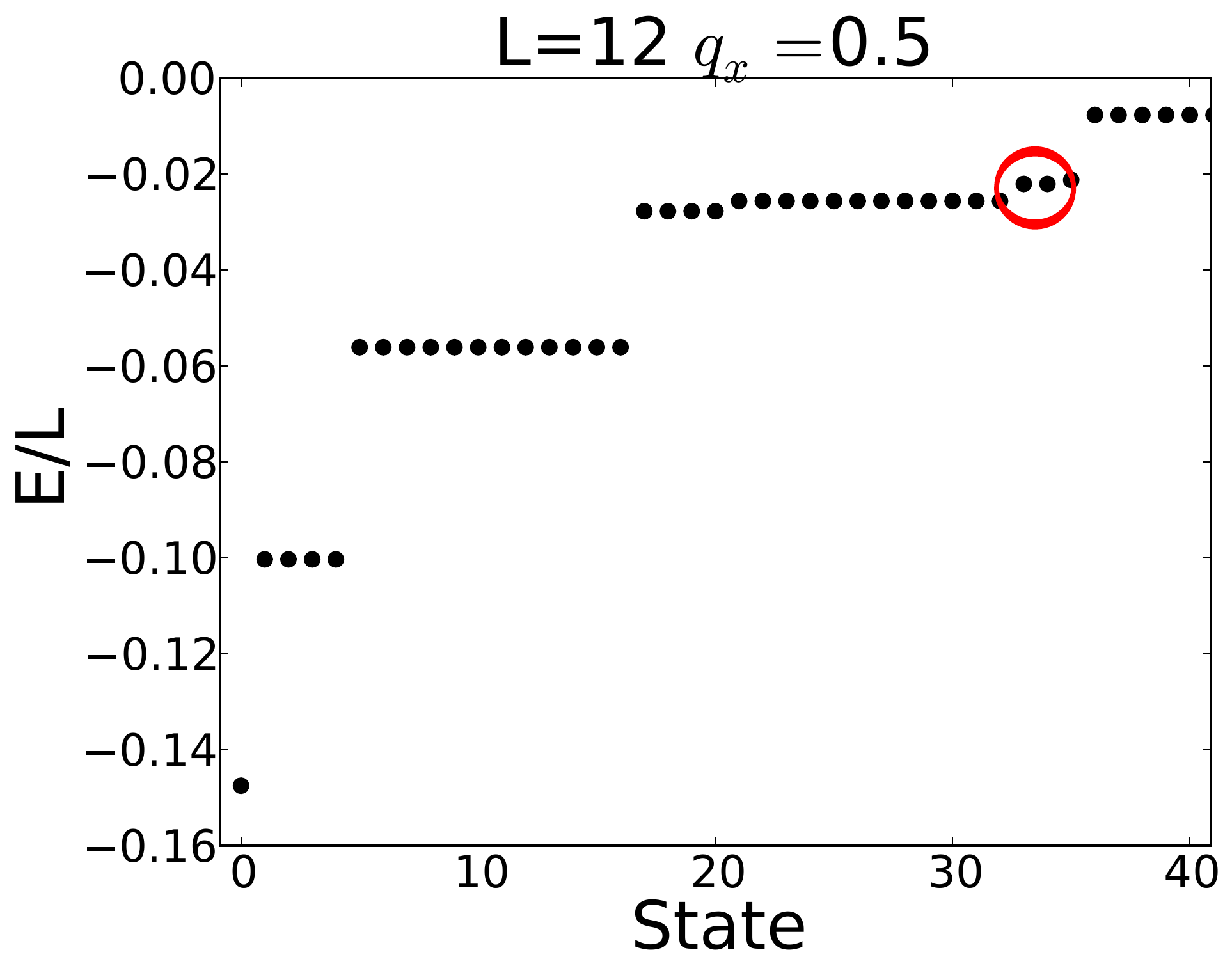}}
\subfigure[]{\includegraphics[width=0.31\linewidth]{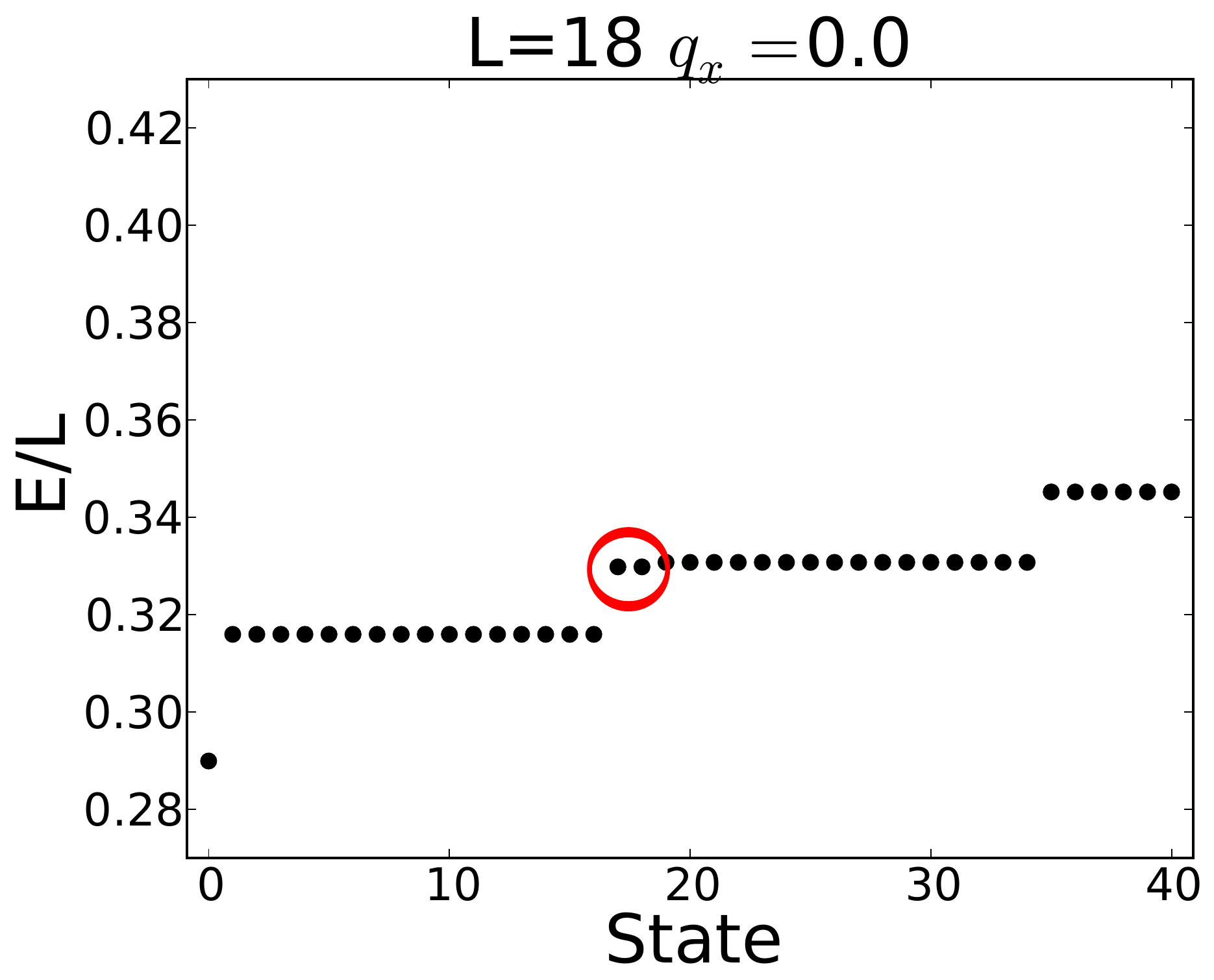}}
\subfigure[]{\includegraphics[width=0.31\linewidth]{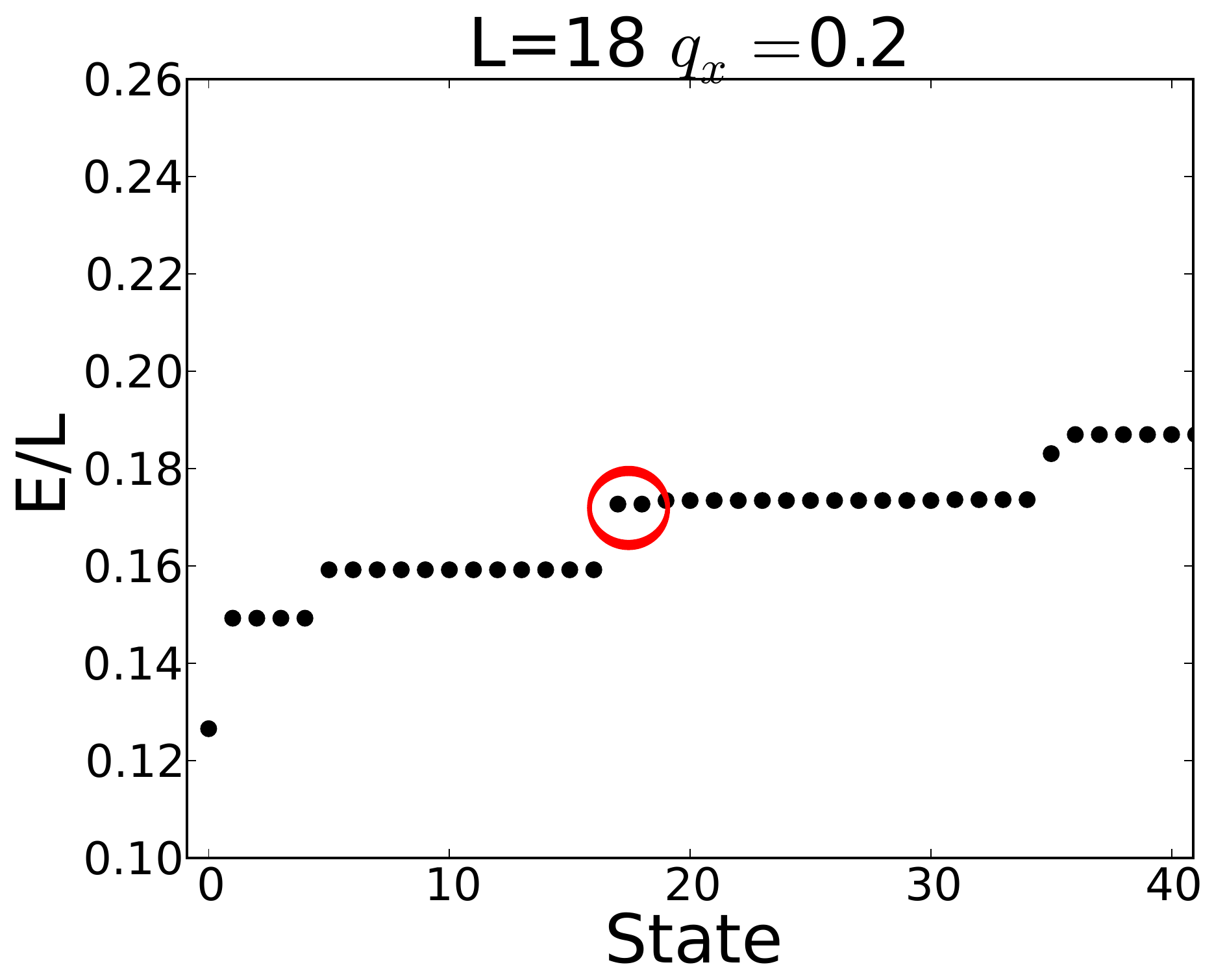}}
\subfigure[]{\includegraphics[width=0.32\linewidth]{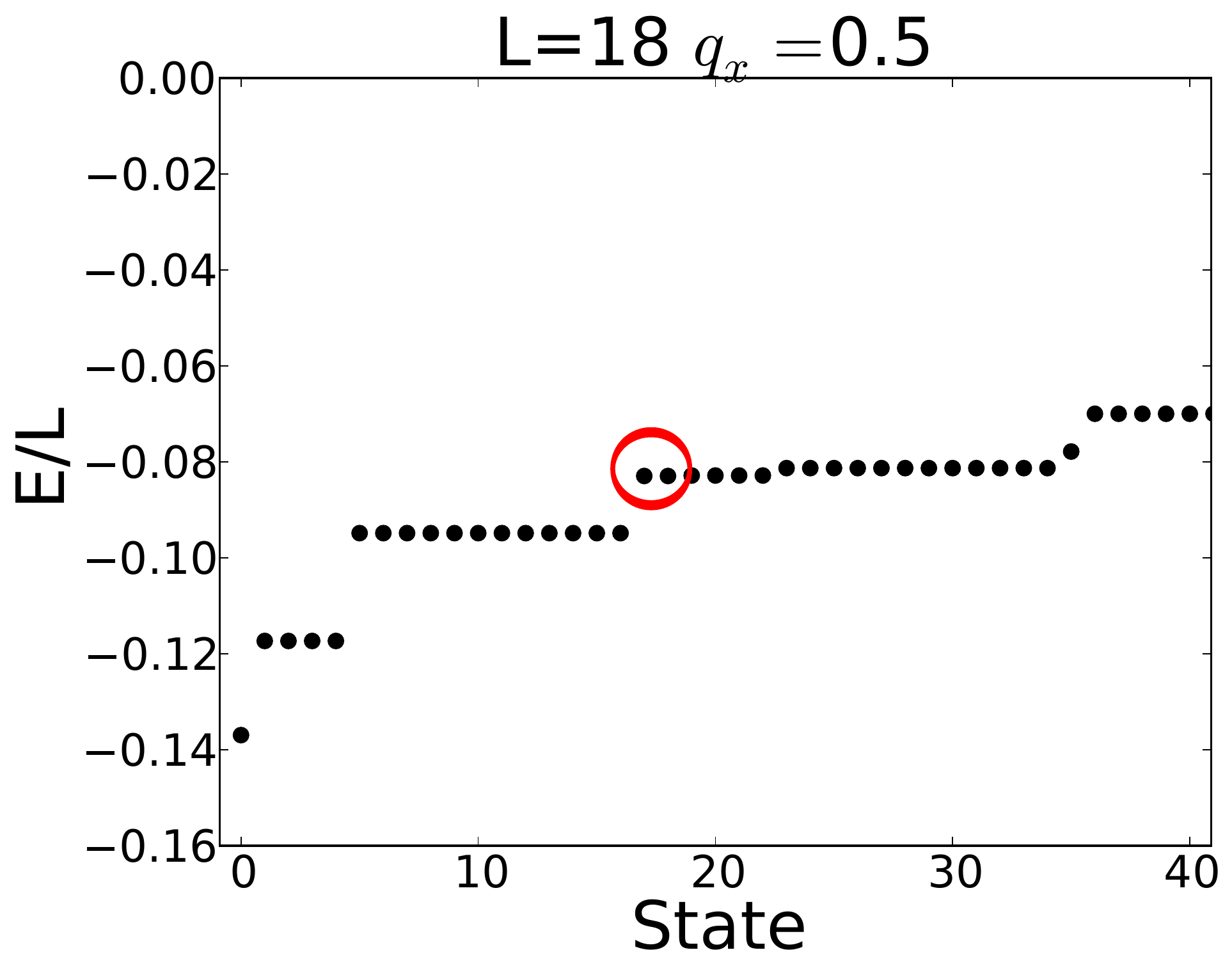}}
\caption{Low energy manifold of the Hamiltonian~\eqref{eq:Hamiltonian} 
for $L=12$ (upper panels) and $L=18$ site \newt{periodic} chains (lower panels) 
for \newt{different} values of $q_x$. For $q_x=0$, shown in panels 
(a),(d) the first excited state is known to be $18$-fold degenerate in the conformal 
limit. On increasing $L$, the inferred trend is that 
the two higher lying singlets (marked by red circles) descend to possibly join the 
$16$-fold exactly degenerate states. (b),(e) and (c),(f) 
show similar trends for $q_x=0.2$ and $q_x=0.5$; in these cases 
the degeneracy structure in the conformal limit 
is narrowed down to a few possibilities.}
\label{fig:degen}
\end{figure*}

For the $12$ site ED results, 
we observe that the low energy manifold consists of a non-degenerate singlet state, 
two sets of $16$-fold degenerate states [\newt{the occurrence of 16 being} 
a consequence of $\mathrm{SU(3)}$ symmetry], 
followed by two degenerate singlets. As can be seen in Fig.~\ref{fig:degen}(a),(d) 
on going from $12$ to $18$ sites, the two singlets descend below the second manifold of $16$ states: 
it is thus conceivable (though not rigorous), that these two states will 
join the $16$-fold degenerate first excited states 
resulting in a $18$-fold degeneracy in the conformal limit.

\begin{figure}[htpb]
\centering
\centering
\includegraphics[width=1\linewidth]{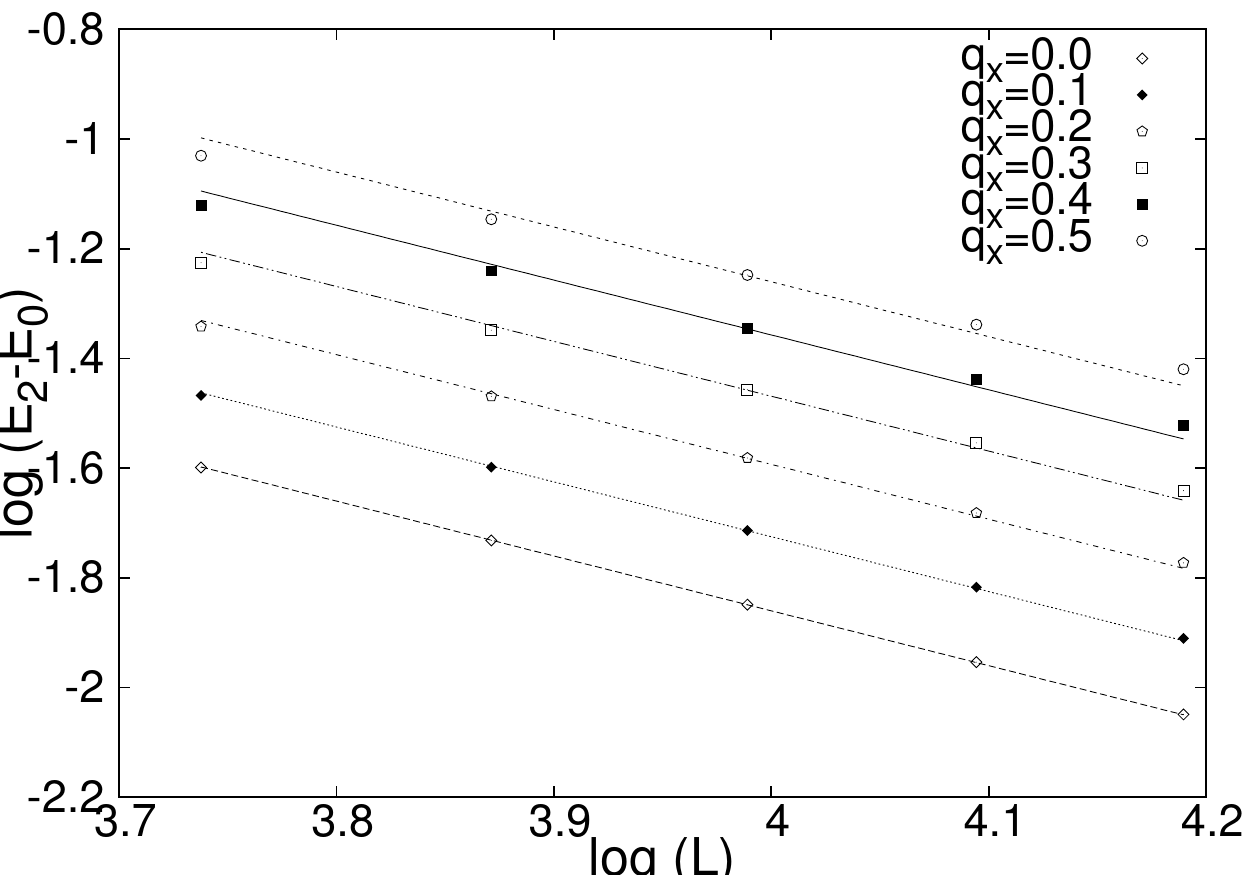}
\caption{Finite size scaling of the second excited state energy gap for various $q_x$. Two 
independent DMRG calculations are performed, one each for the ground state in the $S_z=0$ and $S_z=1$ sectors. 
The fits to Eq.~\eqref{eq:energy_v_scaling_dimension}, along with the knowledge of 
the TLL velocity give the second lowest scaling dimension.}
\label{fig:e2_gap}
\end{figure}

Next, consider the effect of adding the $Q_x$ term with $q_x > 0$. From 
ED, we find that the (exact) $16$-fold degeneracy of 
the first excited state splits; the first excited state is now $4$-fold degenerate, 
all corresponding to $S_z = 0 $ states, and the next excited state is $12$-fold degenerate, 
corresponding to four sets of $S_z = \pm 1 $ and two sets of $S_z = \pm 2$ states. 
Here too, the two degenerate singlets in the low energy spectrum 
descend to lower values on increasing the length of the chain, as can be seen in Fig.~\ref{fig:degen}(b),(e) 
and (c),(f). Based on our experience with the $\mathrm{SU(3)}$ point, we conjecture 
that restoration of conformal symmetry will result in the $4$-fold degeneracy 
being transformed to a $6$-fold degeneracy; although other possibilities are not completely ruled 
out based on this data alone. We expect this
degeneracy structure to hold \newt{on varying $q_x$} only as long as the 
second excited state does not become the third excited state. 

The field theoretic prediction~\eqref{eq:G_qx} confirms these inferences. 
Once the second scaling dimension exceeds the value of 1, 
\newt{ which occurs around $q_x \approx 0.5$}, 
there is a reorganization of \newt{energy degeneracies}. 
\newt{For } $q_x \lesssim 0.5$, we deduce that the quantum numbers 
$\{n_1,n_2,w_1,w_2 \}$ [see Eq.~\ref{internalmomenta}]
corresponding to the lowest $6$ states are 
$\{ \pm1,\pm1,0,0 \}$, $\{ \pm 1,0,0,0 \}$ , $\{ 0,\pm1,0,0 \}$ 
and \newt{those for} the next $12$ states are 
$\{ \mp1,\mp 1,\pm 1,\mp 1 \}$, $\{ \pm 1,0,0, \pm 1 \}$ , $\{ 0,0,\pm 1,0 \}$, 
$\{ 0,0,0,\pm 1 \}$, $\{ 0, \pm 1, \mp 1, 0 \}$ , $\{ 0,0,\pm 1, \mp 1 \}$. 

Figure~\ref{fig:e2_gap} shows fits 
to Eq.~\eqref{eq:energy_v_scaling_dimension}, after taking logarithms of both sides,
to extract the second scaling dimension $x_2$, for various $q_x$; 
similar trends are seen for the first scaling dimension as well. The corrections to 
scaling are found to increase on going from the $q_x =0.0$ to $q_x =0.5$. 
Whether these effects are genuine deviations from the TLL physics or a lack 
of sufficient size to see "true scaling" can not be definitively established within 
our present methodology. \newt{We believe the deviations close to $q_x \approx 0.5$ 
are due to "energy crossings" (i.e. changing multiplet structure), causing additional 
level repulsions.} Thus, one may need very large sizes to get 
precise estimates in this region. 

Despite this source of inaccuracy, the scaling dimensions vary within 10\% 
when they are computed using Eq.~\eqref{eq:energy_v_scaling_dimension} 
for fixed $L$, over the range of lengths considered ($24-66$ sites). 
The obtained values validate the correspondence between the 
lattice model and the CFT and the general trends of their variations with $q_x$ 
support our main conclusions. 

\subsection{Extracting the lowest scaling dimensions from mutual information}
\label{sec:mi}

It is difficult to target multiple excited states in DMRG for long chains, 
especially for a \emph{critical} system where the entanglement entropy 
grows logarithmically with system size. Thus it is extremely desirable 
to have a method to obtain scaling dimensions that involves only the ground state.

\begin{figure}[htpb]
\centering
\subfigure[]{\includegraphics[width=0.8\linewidth]{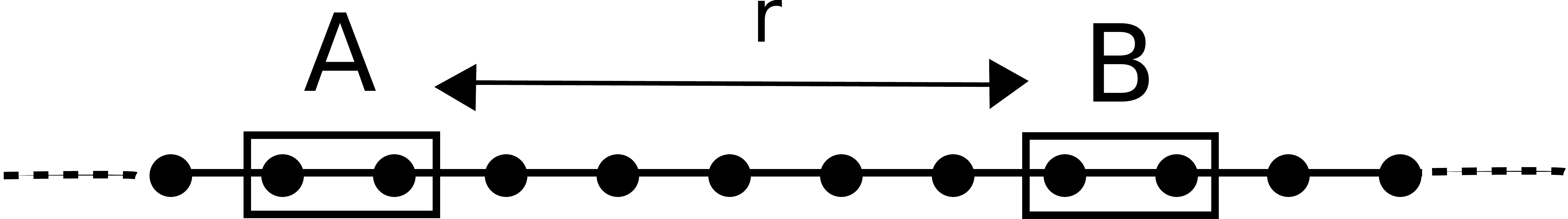} \label{geometries2}}
\subfigure[]{\includegraphics[width=\linewidth]{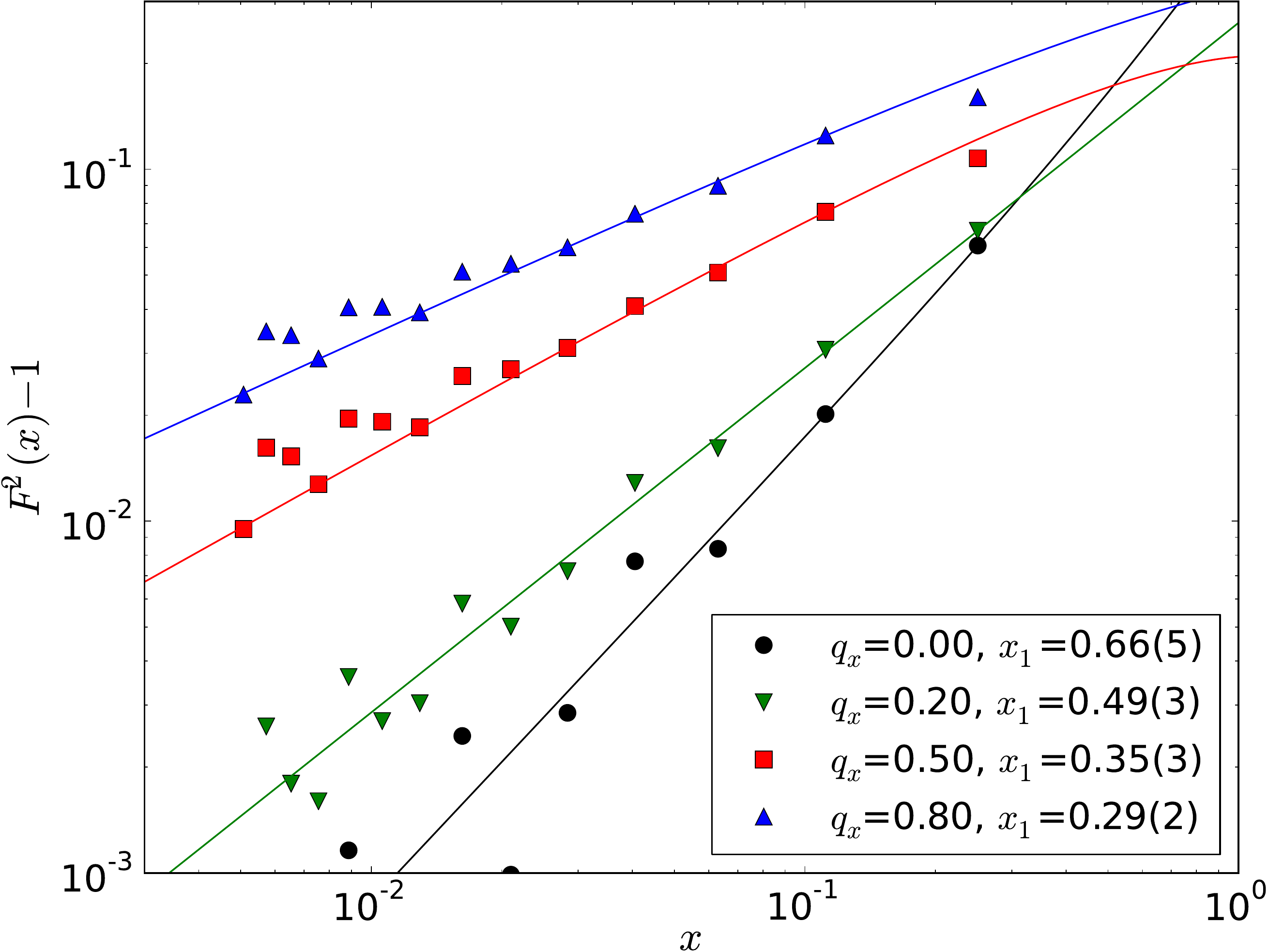} \label{mi_combined}}
\caption{(Color online): (a) Geometry used to compute the mutual information consists of two 
2-site blocks separated by distance $r$. The calculations were performed for a 150-site chain with $r<30$, 
larger $r$ data was discarded to avoid edge effects. Panel (b) shows 
$F^{2}(x)-1$, derived from the mutual information measure Eq.~\eqref{eq:mi} as 
a function of the conformal ratio $x$ (see Eq.~\eqref{eq:conformal_ratio}) 
calculated with DMRG. It was fitted to the analytic form, Eq.~\eqref{eq:Fn} 
to obtain the lowest scaling dimension.}
\label{fig:geometries} 
\end{figure}

Typically this is achieved by measuring ground state correlation functions 
between two distant regions. However, in the most general setting, we a priori 
\emph{do not know} the scaling operators on the lattice i.e. the operators 
whose expectations are to be measured.
To obtain generalized correlation functions between two regions (say $A$ and $B$) 
we calculate their combined reduced density matrix, for varying separations, and extract the 
"mutual information" denoted by $I_{AB}$ and formally defined as,
\begin{equation}
        I_{AB} \equiv S_A + S_B - S_{AUB}, 
\label{eq:mi}
\end{equation}
where $S_A$,$S_B$, $S_{AUB}$ is the EE of regions $A$, $B$ and the union of $A$ and $B$ respectively. 
A schematic of the geometry used for this computation is shown in Fig.~\ref{geometries2}.

The mutual information, unlike the block entanglement entropy, is not directly available in DMRG and 
must be calculated in a matrix product state (MPS) framework.
(Practically, this is achieved by reshaping all left and right optimized transformation matrices 
at the end of the DMRG calculation to get the MPS. Then, the reduced density matrix of 
disjoint regions is calculated using a partial-contraction scheme discussed in Ref.~\cite{Muender}. 
More details of our calculations will be provided elsewhere.)
The mutual information can also be calculated with Monte Carlo methods in 
sign-problem free systems~\cite{Melko_mutual,Wang_Troyer}.

We now discuss extraction of the lowest scaling dimension from $I_{AB}$, for which we briefly 
present known results from the literature. To do so, we closely 
follow Ref.~\cite{Furukawa2009}, whose notations we also use here. 

For a CFT, Calabrese and Cardy (CC)~\cite{Calabrese_Cardy} argued 
that the entanglement entropy of two intervals $A=[x_1, x_2] $ and $B=[x_3,x_4]$ in an infinite lattice 
is given by,
\begin{equation}
	S_{A U B} = \frac{c}{3} \log \left( \frac{x_{21} x_{32} x_{43} x_{41} } {x_{31} x_{42}} \right) + 2s_1, 
\end{equation}
where $x_{ij} \equiv x_i - x_j $. The constant $2s_1$ is determined by demanding that
$S_{AUB} \rightarrow S_A + S_B $ in the limit $x_{21}, x_{43} << x_{31}, x_{42}$. 
Rewriting this formula in terms of the mutual information (i.e. on subtracting out 
the single interval contributions), one gets,
\begin{equation}
	I_{AB}^{CC}= \frac{c}{3} \log \left( \frac{x_{32} x_{41} } {x_{31} x_{42}} \right).  
\end{equation}
For a finite periodic chain, one replaces $x_{ij}$ by the cord distance 
$L/\pi \sin (\pi x_{ij}/L)$, 
this results in,
\begin{equation}
	I_{AB}^{CC}= \frac{c}{3} \log \left( \frac{ \sin(\pi x_{32}/L) \sin(\pi x_{41}/L) } {\sin(\pi x_{31}/L) \sin(\pi x_{42}/L)} \right). 
\end{equation}
It is thus convenient to define the conformal ratio $x$ as,
\begin{equation}
	x \equiv \frac{ \sin(\pi x_{32}/L) \sin(\pi x_{41}/L) } {\sin(\pi x_{31}/L) \sin(\pi x_{42}/L)}. 
\label{eq:conformal_ratio}
\end{equation}
The notion of mutual information can be generalized beyond the von-Neumann entropy, which is 
assigned an index $n=1$, and thus denoted more generally by $I^{(n)}_{AB}$. 
This is achieved by the following replacements in the CC formulae,
\begin{eqnarray}
	S_1  \rightarrow  S_n ,
	\quad 
	c   \rightarrow  \frac{1+n}{6n} c.  
\end{eqnarray}
Ref.~\cite{Furukawa2009} found that the true mutual information and 
the CC mutual information differ by a function $f^{n}(x)$,
\begin{equation}
	I^{(n)}_{AB} - {I^{(n)}}^{CC}_{AB} = f^{(n)}(x), 
\end{equation}
which is reparameterized as, 
\begin{equation}
	\frac{1}{n-1} F^{(n)}(x) \equiv f^{(n)}(x). 
\end{equation}

Calabrese and co-workers~\cite{CCTonni,Alba_Calabrese} have shown that for $n>1$ and in the limit of small $x$,
\begin{align}
	F^{(n)}(x)-1&= \left ( \frac{x}{4n^{2}} \right )^{\alpha} s_{2}(n)+ \left ( \frac{x}{4n^{2}} \right )^{2 \alpha} s_{4}(n) 
	\nonumber \\
	& \quad +(\text{higher order}), 
\label{eq:Fn}
\end{align}
where $\alpha$ is twice the lowest scaling dimension $x_1$. 
The coefficients $s_2(n)$ and $s_4(n)$ are the contributions in the small $x$ 
expansion coming from the two and four-point functions of the operator in the CFT with the lowest scaling dimension.
 
Two concerns when using equation~\eqref{eq:Fn} in numerical simulations 
are (1) it holds only for an infinite lattice and (2) 
it assumes that the non-zero contributions are solely from the operator 
with the \emph{lowest} scaling dimension. However, for a 
finite system there are contributions from \emph{all} operators. 
Thus the lowest scaling dimension fitted is simply an effective one 
trying to mimic the action of a linear combination of many (different scaling) operators. 
Empirically, for an open chain of 150 sites, all the errors (systematic and due to fitting) 
appear to be within 10\%, which is roughly the error
we also obtain from fitting to energies. 

Our results for fits to a power law for $F^{2}(x)-1$ for various $q_x$ 
are shown in Fig.~\ref{mi_combined}. The overall fits are reasonable, 
though there are local features not captured by Eq.~\eqref{eq:Fn}: just like the case of the EE, 
these are attributed to open boundaries. Such features are also seen 
in the spin-1/2 XXZ model, studied independently by Ref.~\cite{Barcza}.

\begin{figure}[htpb]
\centering
\centering
\includegraphics[width=1\linewidth]{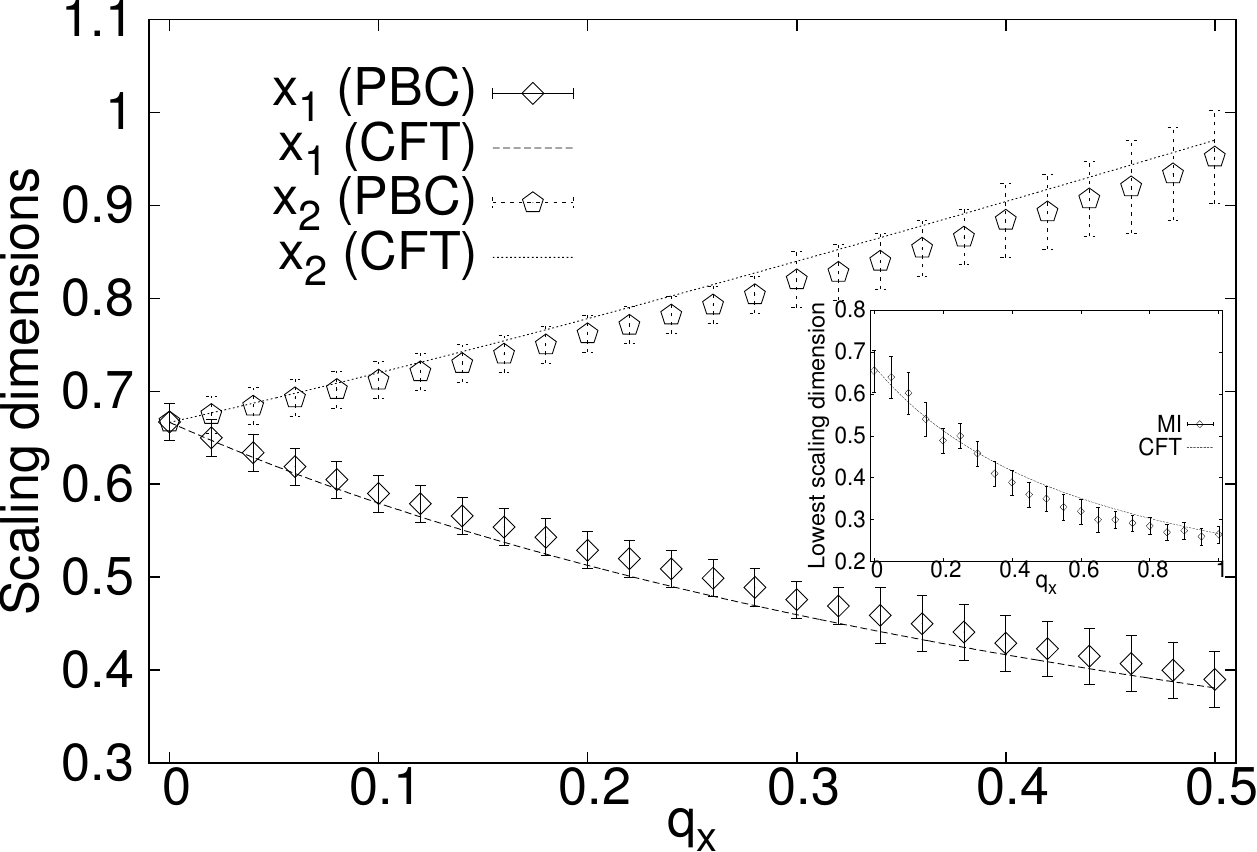}
\caption{The main panel shows the first 
two scaling dimensions, $x_1$ and $x_2$, 
as a function of $q_x$, obtained from finite size scaling of energy gaps 
obtained from a combination of exact diagonalization (ED) 
and the density matrix renormalization group (DMRG) for periodic 
chains denoted by PBC. The $c=2$ conformal 
field theory (CFT) prediction is also shown. 
The inset shows the lowest scaling dimension obtained from 
the mutual information (MI) measure (see text), computed within the 
DMRG/Matrix product state framework for an open chain of 150 sites.}
\label{fig:two_sd} 
\end{figure}	

\subsection{Extraction of TLL parameters}
\label{sec:LL}

Fig.~\ref{fig:two_sd} shows the lowest two scaling dimensions 
obtained from finite size scaling of energy gaps as a function of $q_x$. 
The inset shows the lowest scaling dimension 
from the mutual information method; with this metric, 
we were able to explore a larger range of $q_x$.
The general agreement (within errors) between these independent metrics 
confirms our that we can reliably calculate lowest scaling dimensions. 
Thus we proceed to discuss the extraction method for the four TLL parameters. 

Given a trial set of $G_{ab}$ and $B_{ab}$, we calculate the 
lowest 18 scaling dimensions, which need not be distinct, 
and denote them by $x_i^{\text{G,B}}$. 
We then evaluate a cost function,
\begin{equation}
	C\text{(G,B)} \equiv \sum_{i} (x_i^{\text{G,B}} - x_i^{\text{DMRG}})^2
\end{equation}
and minimize it with respect to $G_{11},G_{12},G_{22}$ and $B_{12}$ 
to obtain the best fit. We used the Nelder-Mead simplex algorithm 
built into the GNU Scientific library for this purpose. 

In order to confirm our inferences about the nature of the degeneracies in the low energy manifold, 
we attempted to fit to two degeneracy structures for the 
first and second excited states. First, we assumed that the degeneracy (denoted by $g_i$) 
of the first two distinct scaling dimensions to be $(g_1,g_2)=(6,12)$ and in the 
second case $(g_1,g_2)=(4,12)$. In all cases, 
for $q_x<0.5$, we found the former gave a significantly better fit to the CFT 
formulae~\eqref{eq:scaling_dimensions}. In fact, attempts to use the $(4,12)$ structure 
gave optimized solutions closer to a $(g_1,g_2,g_3)=(4,2,12)$ degeneracy structure, 
hinting that the imposed structure was incorrect. The quality of our fits are 
checked by how well the scaling dimensions were reproduced; for the correct degeneracy structure, these agreed to within $\pm 0.03$. 

\begin{figure}[htpb]
\centering
\centering
\includegraphics[width=1\linewidth]{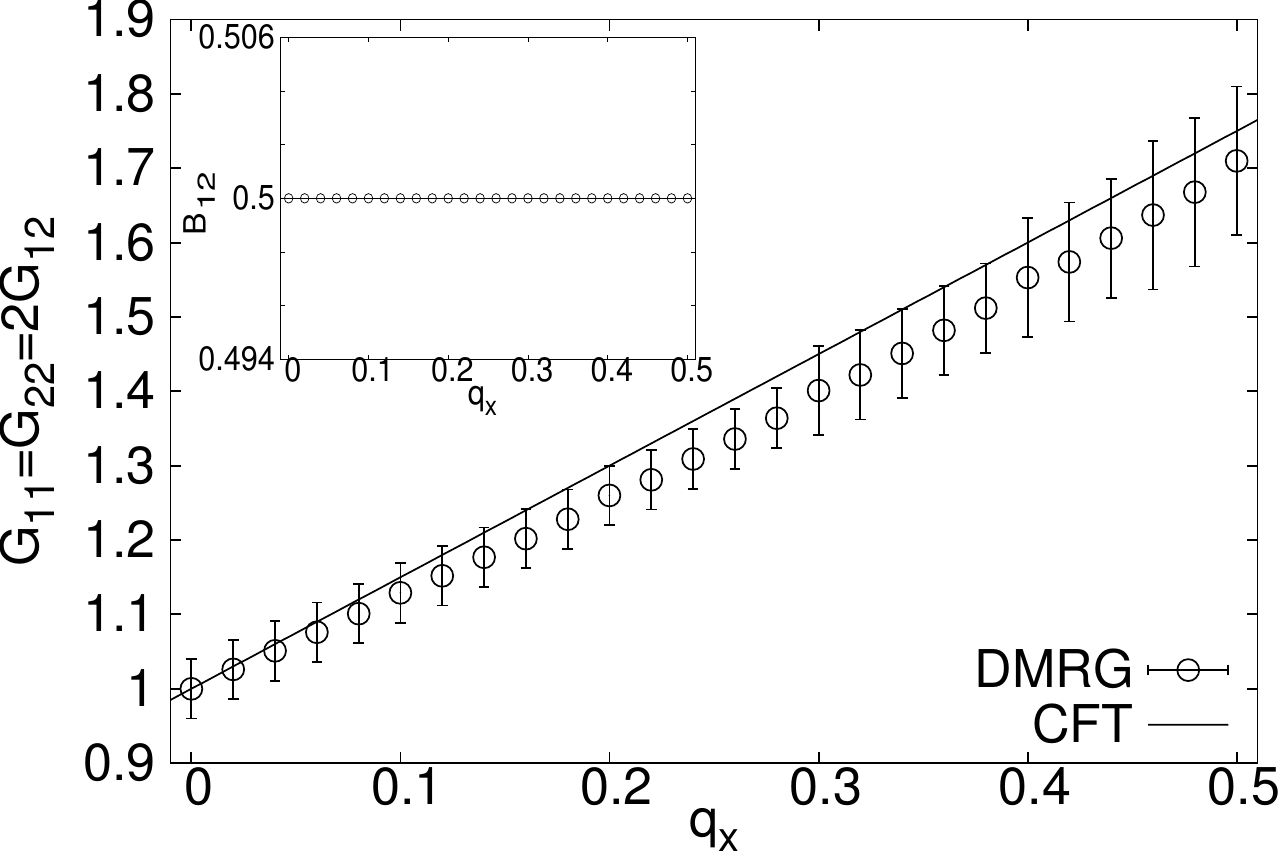}
\caption{TLL parameters as a function of $q_x$ 
extracted from matching scaling dimensions to a $c=2$ CFT. The main panel 
shows $G_{11}$ as a function of $q_x$ and we find $G_{11}=G_{22}=2G_{12}$.
The inset shows $B_{12}$ which is found to be constant. 
The CFT prediction is shown by the solid lines.}
\label{fig:GB_Qx} 
\end{figure}

The agreement of the values of the measured and expected scaling dimensions, shown in 
Fig.~\ref{fig:two_sd}, strongly indicates an internally consistent 
scenario for the lattice model to CFT mapping. 
This is also equivalently seen in the extracted TLL parameters, 
shown in Fig.~\ref{fig:GB_Qx}, which are consistent with Eq.~\eqref{eq:G_qx}: 
they satisfy the expected relation $G_{11}=G_{22}=2 G_{12}$. 
The relative error in the scaling dimensions propagates to these parameters;
for example, the overall error in $G_{11}$ is roughly twice the error in $x_2$.
As expected from the duality explained in section~\ref{sec:general_boson}, 
the scaling dimensions depend on $B_{12}$ up to an integer shift. 
Thus we focus on a particular representation and find that $B_{12}=1/2$ explains our 
data for all $q_x$. Finally, even though we have shown data only for $q_x<0.5$, 
the mutual information data in Fig.~\ref{fig:two_sd} (inset) 
suggests the validity of the theory for larger $q_x$. 

\section{Conclusion} 
\label{sec:conclusion}
In conclusion, we have developed an analytic 
correspondence between $c=2$ free boson theories 
and microscopic spin-1 models, using bosonization techniques. 
For the particular form of Hamiltonian considered (Eq.~\eqref{eq:Hamiltonian}), 
we made a prediction for the value of the Tomonaga-Luttinger liquid (TLL) 
parameters as a function of $q_x$, the parameter characterizing the lattice model. 

To build evidence on the numerical front, 
we performed exact diagonalization (ED) and density matrix renormalization group (DMRG) 
calculations to obtain the lowest scaling dimensions from the energetics of the system; 
a scheme feasible for short periodic chains. 
However, our use of the mutual information entropy between disjoint blocks, 
calculated solely from the ground state, provides a promising route to extend 
these calculations for long chains. 

Using this numerical data and the mapping from spin-chains to $c=2$ theories, 
we deduced the value of all four TLL parameters as a function of $q_x$. We expect 
our analyses to apply to more general situations, for eg. the model in Eq.~\eqref{eq:Hamiltonian} 
with non zero $q_y$. In future work, we aim to extend these ideas 
to calculate multiple low-lying scaling operators and dimensions 
of the $c=2$ CFT using the correlation density matrix~\cite{Cheong_Henley,Muender}. 

Our broader objective is an effort to develop generic 
methods to map lattice models to multi-component field theories. 
We anticipate that this multi-scale modelling approach will be useful for understanding 
the physics at very large length scales; sizes 
that may not be directly accessible in numerical simulations. 
Once we have built confidence in the mapping between 
the lattice and continuum descriptions, we can use the (often known) 
predictions of the latter.

\section{Acknowledgements} 
We thank Christopher Henley, Dunghai Lee, Eduardo Fradkin, Shunsuke Furukawa, Garnet Chan, 
Andreas L\"auchli, Victor Chua and Norman Tubman for discussions. 
This work has been supported by SciDAC grant DE-FG02-12ER46875. 
Computer time was provided by XSEDE and the Taub campus cluster at the University of 
Illinois Urbana-Champaign/NCSA. SR is supported by the A. P. Sloan Foundation. 

\bibliographystyle{prsty}
\bibliography{refs}

\begin{thebibliography}{10}

\bibitem{Calabrese_Cardy}
P. Calabrese and J. Cardy, Journal of Statistical Mechanics: Theory and
  Experiment {\bf 2004},  P06002  (2004).

\bibitem{Li_Haldane}
H. Li and F.~D.~M. Haldane, Phys. Rev. Lett. {\bf 101},  010504  (2008).

\bibitem{Ryu}
S. Ryu and T. Takayanagi, Phys. Rev. Lett. {\bf 96},  181602  (2006).

\bibitem{Headrick}
M. Headrick, Phys. Rev. D {\bf 82},  126010  (2010).

\bibitem{Tubman}
J. McMinis and N. Tubman, Phys. Rev. B {\bf 87},  081108  (2013).

\bibitem{Melko}
M. Hastings, I. Gonz\'alez, A. Kallin, and R. Melko, Phys. Rev. Lett. {\bf
  104},  157201  (2010).

\bibitem{Ryu_Hatsugai}
S. Ryu and Y. Hatsugai, Phys. Rev. B {\bf 73},  245115  (2006).

\bibitem{Thomale2010}
R. Thomale, D.~P. Arovas, and B.~A. Bernevig, Phys. Rev. Lett. {\bf 105},
  116805  (2010).

\bibitem{Lundgren2014}
R. Lundgren {\it et~al.}, Phys. Rev. Lett. {\bf 113},  256404  (2014).

\bibitem{Cheong_Henley}
S.-A. Cheong and C. Henley, Phys. Rev. B {\bf 79},  212402  (2009).

\bibitem{Muender}
W. Muender {\it et~al.}, New Journal of Physics {\bf 12},  075027  (2010).

\bibitem{Henley_Changlani}
C.~L. Henley and H.~J. Changlani, Journal of Statistical Mechanics: Theory and
  Experiment {\bf 2014},  P11002  (2014).

\bibitem{Melko_mutual}
R. Melko, A. Kallin, and M. Hastings, Phys. Rev. B {\bf 82},  100409  (2010).

\bibitem{Barcza}
G. Barcza, R. M. Noack, J. Solyom, O. Legeza, arXiv:1406.6643 (unpublished).

\bibitem{Furukawa2009}
S. Furukawa, V. Pasquier, and J. Shiraishi, Phys. Rev. Lett. {\bf 102},  170602
   (2009).

\bibitem{Tomonaga}
S.-i. Tomonaga, Progress of Theoretical Physics {\bf 5},  544  (1950).

\bibitem{Luttinger_original}
J.~M. Luttinger, Journal of Mathematical Physics {\bf 4},    (1963).

\bibitem{Giamarchi}
T. Giamarchi, Quantum Physics in One Dimension, Volume 121 of International
  Series of Monographs on Physics, Clarendon Press, (2003).

\bibitem{Haldane_Luttinger}
F.~D.~M. Haldane, Journal of Physics C: Solid State Physics {\bf 14},  2585
  (1981).

\bibitem{Bockrath}
M. Bockrath, D. H. Cobden, J. Lu, A. G. Rinzler, R. E. Smalley, L. Balents and
  P. L. McEuen, \emph{Nature} 397, 598-601 (1999).

\bibitem{Ishii}
H. Ishii, H. Kataura, H. Shiozawa, H. Yoshioka, H. Otsubo, Y. Takayama, T.
  Miyahara, S. Suzuki, Y. Achiba, M. Nakatake, T. Narimura, M. Higashiguchi, K.
  Shimada, H. Namatame, M. Taniguchi, \emph{Nature} 426, 540-544 (2003).

\bibitem{Yacoby}
A. Yacoby {\it et~al.}, Phys. Rev. Lett. {\bf 77},  4612  (1996).

\bibitem{dmrg_white}
S.~R. White, Phys. Rev. Lett. {\bf 69},  2863  (1992).

\bibitem{Lauchli_operator}
A.M. L\"auchli,arxiv:1303.0741 (unpublished).

\bibitem{Jeckelmann_DMRG}
E. Jeckelmann, Journal of Physics: Condensed Matter {\bf 25},  014002  (2013).

\bibitem{Karrasch}
C. Karrasch and J. Moore, Phys. Rev. B {\bf 86},  155156  (2012).

\bibitem{Dalmonte}
M. Dalmonte, E. Ercolessi, and L. Taddia, Phys. Rev. B {\bf 85},  165112
  (2012).

\bibitem{Rex_TLL}
R. Lundgren, Y. Fuji, S. Furukawa, and M. Oshikawa, Phys. Rev. B {\bf 88},
  245137  (2013).

\bibitem{Sakai}
T. Sakai {\it et~al.}, Journal of Physics: Condensed Matter {\bf 22},  403201
  (2010).

\bibitem{Schulz1}
H.~J. Schulz, Phys. Rev. B {\bf 53},  R2959  (1996).

\bibitem{Schulz2}
H.J. Schulz, ``Correlated Fermions and Transport in Mesoscopic Systems'', ed.
  T. Martin, G. Montambaux, J. Tran Thanh Van (Editions Frontieres,
  Gif--sur--Yvette, 1996), p. 81.

\bibitem{Schnack}
J. Schnack {\it et~al.}, Phys. Rev. B {\bf 70},  174420  (2004).

\bibitem{Manaka}
H. Manaka {\it et~al.}, Journal of the Physical Society of Japan {\bf 78},
  093701  (2009).

\bibitem{Bose_metal}
R.~V. Mishmash {\it et~al.}, Phys. Rev. B {\bf 84},  245127  (2011).

\bibitem{Balents_spin_tube}
R. Chen {\it et~al.}, Phys. Rev. B {\bf 87},  165123  (2013).

\bibitem{Sato_PRB}
M. Sato, Phys. Rev. B {\bf 75},  174407  (2007).

\bibitem{Qin_torus}
M.~P. Qin {\it et~al.}, Phys. Rev. B {\bf 86},  134430  (2012).

\bibitem{Lai}
C.~K. Lai, Journal of Mathematical Physics {\bf 15},    (1974).

\bibitem{Sutherland}
B. Sutherland, Phys. Rev. B {\bf 12},  3795  (1975).

\bibitem{Itoi_Kato}
C. Itoi and M.-H. Kato, Phys. Rev. B {\bf 55},  8295  (1997).

\bibitem{Papa}
N. Papanicolaou, Nuclear Physics B {\bf 305},  367   (1988).

\bibitem{Haldane}
F.~D.~M. Haldane, Phys. Rev. Lett. {\bf 50},  1153  (1983).

\bibitem{AKLT}
I. Affleck, T. Kennedy, E.~H. Lieb, and H. Tasaki, Phys. Rev. Lett. {\bf 59},
  799  (1987).

\bibitem{Millet_expt}
P. Millet {\it et~al.}, Phys. Rev. Lett. {\bf 83},  4176  (1999).

\bibitem{Oshikawa_TLL}
M. Oshikawa, C. Chamon, and I. Affleck, Journal of Statistical Mechanics:
  Theory and Experiment {\bf 2006},  P02008  (2006).

\bibitem{Affleck_Les}
I. Affleck, in Fields, Strings and Critical Phenomena, 1988 Les Houches Lecture
  Notes, edited by E. Brezin and J. Zinn-Justin (Elsevier, Amsterdam, 1989), p.
  564.

\bibitem{yellowbook}
P.~D. Francesco, P. Mathieu, and D. Senechal, {\em Conformal Field Theory},
  {\em Graduate Texts in Contemporary Physics} (Springer, ADDRESS, 1997).

\bibitem{GSW}
M. Green, J. Schwarz, and E. Witten, {\em Superstring Theory: Introduction},
  {\em Cambridge monographs on mathematical physics} (Cambridge University
  Press, ADDRESS, 2012).

\bibitem{Ginsparg_Les}
P. Ginsparg, in Fields, Strings and Critical Phenomena: Proceedings (Les
  Houches 1988), ed. by E. Brezin and Jean Zinn-Justin, pp. 1-168. Amsterdam:
  North-Holland (1990).

\bibitem{Dulat}
S. Dulat and K. Wendland, Journal of High Energy Physics {\bf 2000},  012
  (2000).

\bibitem{becker2006string}
K. Becker, M. Becker, and J. Schwarz, {\em String Theory and M-Theory: A Modern
  Introduction} (Cambridge University Press, ADDRESS, 2006).

\bibitem{Giveon}
A. Giveon, M. Porrati, and E. Rabinovici, Physics Reports {\bf 244},  77
  (1994).

\bibitem{ALPS}
B. Bauer {\it et~al.}, Journal of Statistical Mechanics: Theory and Experiment
  {\bf 2011},  P05001  (2011).

\bibitem{Itoi_Affleck}
C. Itoi, S. Qin, and I. Affleck, Phys. Rev. B {\bf 61},  6747  (2000).

\bibitem{Fath}
G. F\'ath and J. S\'olyom, Phys. Rev. B {\bf 51},  3620  (1995).

\bibitem{Rachel_SU3}
M. F{\"u}hringer {\it et~al.}, Annalen der Physik {\bf 17},  922  (2008).

\bibitem{Aguado_SU3}
M. Aguado {\it et~al.}, Phys. Rev. B {\bf 79},  012408  (2009).

\bibitem{Lauchli_Schmid_SU3}
A. L\"auchli, G. Schmid, and S. Trebst, Phys. Rev. B {\bf 74},  144426  (2006).

\bibitem{Note1}
This observation can possibly be better understood by extracting the scaling
  operators from numerics. This involves determining a coarse grained operator
  that spans three sites; a direction we will not explore in the present paper.

\bibitem{Note2}
Practically this was checked for $0 \leq q_x<100$.

\bibitem{Wang_Troyer}
L. Wang and M. Troyer, Phys. Rev. Lett. {\bf 113},  110401  (2014).

\bibitem{CCTonni}
P. Calabrese, J. Cardy, and E. Tonni, Journal of Statistical Mechanics: Theory
  and Experiment {\bf 2011},  P01021  (2011).

\bibitem{Alba_Calabrese}
V. Alba, L. Tagliacozzo, and P. Calabrese, Phys. Rev. B {\bf 81},  060411
  (2010).

\end{thebibliography}

\end{document}